\documentclass[reprint,amsmath,amssymb,aps,superscriptaddress,prb]{revtex4-2}

\usepackage{graphicx}
\usepackage[pdftex]{color}
\usepackage{dcolumn}

\usepackage{bm}
\usepackage{color}
\usepackage{comment}

\usepackage{braket}

\usepackage{here}

\usepackage{appendix}

\newcommand{\up}{\uparrow}
\newcommand{\down}{\downarrow}

\begin{document}

\preprint{APS/123-QED}

\title{Nonlinear response induced by Ferromagnetism in a Noncentrosymmetric Kondo Lattice system}

\author{Koki Shinada}
\email{shinada.koki.64w@st.kyoto-u.ac.jp}
\author{Robert Peters}
\affiliation{Department of Physics, Kyoto University, Kyoto 606-8502, Japan}

\date{\today}

\begin{abstract}
Recently, nonlinear responses have been actively studied in both experiments and theory. Particularly interesting are inversion-symmetry broken systems, where an even-order nonlinear electrical conductivity can be nonzero, resulting in nonreciprocity. Second-order nonlinear conductivities attract much attention because of their sensitivity to detect inversion-symmetry breaking in materials and their functionalities.
However, while the nonlinear response has been actively studied in noninteracting systems for a long time, the nonlinear response in strongly correlated materials is still poorly understood. This paper analyzes the nonlinear conductivity in a correlated noncentrosymmetric system, namely a Kondo lattice system with Rashba type spin-orbit coupling. We mainly focus on the ferromagnetic phase, in which the second-order nonlinear conductivity becomes finite. Remarkably, we find that the second-order conductivity becomes only finite perpendicular to the ferromagnetic magnetization and has a strong spin dependence; due to a gap at the Fermi energy for one spin direction, the linear and nonlinear conductivity is only finite for the ungapped spin direction.
Finally, we analyze sign changes in the nonlinear conductivity, which can be explained by a combination of correlation effects and the energetic shift due to the occurring ferromagnetism. 
\end{abstract}

\maketitle

\section{Introduction} 
Systems with strong spin-orbit coupling (SOC) have attracted broad interest in condensed matter physics, atomic physics, and others due to various exciting phenomena \cite{annurev-conmatphys-020911-125138,Manchon2015}. SOC is a relativistic effect that is usually small in solids. However, in $f$-electron systems, which include heavy elements, SOC can become strong. Recently, SOC in inversion-symmetry broken systems has been energetically studied in condensed matter physics, where the antisymmetric SOC (ASOC), e.g., the Rashba-type SOC, can appear \cite{Manchon2015}. Besides intrinsic noncentrosymmetric systems \cite{Smidman_2017}, ASOC appears at interfaces of different two-dimensional (2D) materials, where the inversion symmetry is broken perpendicular to the interface \cite{PhysRevLett.109.157006,PhysRevLett.112.156404,Lin2019}. ASOC locks the spin of the electrons at each momentum, making it possible to manipulate the spin of the electrons by applying an electric field or electric current\cite{PhysRevB.97.115128,Chernyshov2009}, which can be used in spintronics devices. 
For example, an electric field or electric current can induce a magnetization in materials with ASOC, which is known as the magnetoelectric effect or Edelstein effect \cite{EDELSTEIN1990233}. ASOC also plays a vital role in topological materials \cite{Hasan2010} because ASOC hybridizes orbitals with different parity in the Brillouin zone except for high symmetry momenta, resulting in a nontrivial band structure.

Besides in spintronics applications, inversion symmetry breaking plays an essential role in the observation of nonlinear responses.
DC nonlinear responses have been studied for decades such as the electric magnetochiral anisotropy which is observed in polar systems  \cite{Rikken2001,Ishizaka2011,Rikken2015,Avci2015,Avci20152,Olejn2015,Ideue2017,Yasuda2016}, chiral systems \cite{Krstic2002,Pop2014,Yokouchi2017,Aoki2019,Kitaori2021}, Weyl semimetals \cite{Morimoto20162},  superconductors \cite{Wakatsukie1602390}, and has been analyzed theoretically in Ref. \cite{Ishizuka2020}.  
Furthermore, nonlinear second-order optical responses, including the shift current and second harmonic generation (SHG), have attracted great interest because they
provide rich information about the band structure, especially the geometrical aspect. The shift current is a DC current proportional to $E(\omega)E(-\omega)$, related to the difference in the Berry connection between different bands \cite{Sipe2000,Morimoto2016,Nagaosa2017}. SHG is an AC current where the frequency is $2\omega$ proportional to $E(\omega)E(\omega)$. A large SHG is observed, for example, in the Weyl semimetal, TaAs \cite{Wu2017}. 
In addition, nonlinear optical responses related to the Berry curvature \cite{Moore2010}, or the Berry curvature dipole \cite{Sodemann2015} have been discussed. 

An intriguing example of a DC nonlinear response appearing in inversion symmetry broken materials is the nonreciprocal current \cite{Tokura2018}, where the absolute value of the current depends on the direction of the electric field. This is a well-known phenomenon in p-n junctions but can also be realized in bulk systems when the inversion symmetry is broken. 
If the current $J$ is expanded in the DC electric field as $E$, $J=\sigma^{(1)} E + \sigma^{(2)} E^2 + \cdots$, a finite second-order conductivity, $\sigma^{(2)}$ can appear in noncentrosymmetric systems resulting in nonreciprocity.

Nonlinear responses are mainly studied in noninteracting systems but are less well understood in interacting systems.  Remarkably, a giant nonlinear Hall effect has been recently observed in a strongly correlated material, $\mathrm{Ce_3Bi_4Pd_3}$ \cite{Dzsaber2021}. The magnitude of the observed nonlinear Hall effect has been explained by the renormalization effect due to strong correlations \cite{Michishita2021,Kofuji2021}. 
These experimental results and subsequent theoretical calculations show that nonlinear responses can be strongly enhanced in correlated materials. Thus, noncentrosymmetric correlated materials appear as an exciting platform to study nonlinear responses experimentally and theoretically. 

Motivated by these previous studies observing a large nonlinear response in strongly correlated materials,  we focus in this study on other correlation-induced phenomena, namely, magnetism and the Kondo effect.
We examine magnetism and its impact on the nonreciprocal response in a noncentrosymmetric $f$ electron system; in particular, we study ferromagnetism in a Kondo lattice system with Rashba-type SOC. 
First, we analytically derive the behavior of the nonlinear conductivity induced by the ferromagnetic state using the Boltzmann equation. We find that the nonlinear conductivity is induced by the Rashba-type SOC combined with ferromagnetism; a ferromagnetic magnetization is essential to observe it. Furthermore, the nonlinear current is perpendicular to the ferromagnetic magnetization. Second, we microscopically analyze ferromagnetism, the linear and nonlinear conductivity using dynamical mean-field theory (DMFT) \cite{Metzner1989,Georges1996,Pruschke1995}. We obtain the ferromagnetic phase diagram and the dependence of the conductivity on the temperature and magnetization. 
We show that the density of states has a gap for the majority spin direction in the ferromagnetic state.  Thus, the conductivity strongly depends on the spin direction. In addition, we find that the nonlinear conductivity exhibits sign changes depending on the temperature and the interaction strength. One sign change is driven by the magnetic energy shift and the Kondo effect. The other sign change is induced by the changes of the self-energy at finite temperature.

The rest of the paper is organized as follows: In Sec.~\ref{model}, we introduce our model Hamiltonian, namely the periodic Anderson model with a Rashba-type SOC. In Sec.~\ref{Boltzmann}, we derive the Kondo Lattice model with a Dzyaloshinskii-Moriya interaction and calculate the linear and second-order conductivity using the Boltzmann equation. In Sec.~\ref{Micro}, we microscopically analyze the ferromagnetic phase and the conductivity using DMFT. Finally, In Sec.~\ref{conclusion}, we summarize our results.

\section{Model} \label{model}
We use a two-dimensional periodic Anderson model to analyze the effect of magnetism of localized electrons on the conductivity of itinerant electrons. We introduce a Rashba-type spin-orbit coupling(SOC) originating from an inversion symmetry breaking, which is necessary to observe a  nonreciprocal current. Although we here use a two-dimensional system for simplicity, we believe that our results remain valid for three-dimensional noncentrosymmetric systems. 
Thus, our model can be understood as a simplified version of a 3D noncentrosymmetric $f$ electron material\cite{JPSJ.75.043703,PhysRevLett.95.247004,PhysRevLett.92.027003,JPSJ.84.074702,PhysRevB.99.155141} or a single layer of the recently studied $f$ electron superlattices\cite{PhysRevLett.109.157006,PhysRevLett.112.156404}, where the inversion symmetry is broken at the interface between different materials.
Our model Hamiltonian reads
\begin{eqnarray}
&&H = H_c + H_f  + H_{cf1} + H_{cf2} \\
&&H_c=\sum_{\bm{k} \sigma} \varepsilon_{\bm{k} \sigma} c^{\dagger}_{\bm{k} \sigma} c_{\bm{k} \sigma} \\
&&H_f = \varepsilon_f \sum_{j \sigma} f^{\dagger}_{j \sigma} f_{j \sigma} + U \sum_{j} n_{f j \up} n_{f j \down} \\
&& H_{cf1} = V \sum_{\bm{k} \sigma} (c^{\dagger}_{\bm{k} \sigma} f_{\bm{k} \sigma} + f^{\dagger}_{\bm{k} \sigma} c_{\bm{k} \sigma}) \\
&& H_{cf2} = V' \sum_{\bm{k} \sigma \sigma'} ((i \bm{d}_{\bm{k}} \cdot \bm{\sigma}_{\sigma \sigma'}) c^{\dagger}_{\bm{k} \sigma} f_{\bm{k} \sigma'} +(-i \bm{d}^*_{\bm{k}} \cdot \bm{\sigma}_{\sigma \sigma'}) f^{\dagger}_{\bm{k} \sigma} c_{\bm{k} \sigma'}) \nonumber \\ \\
&&\bm{d}_{\bm{k}} = -i (\sin k_y , - \sin k_x ),
\end{eqnarray}
where $c^{(\dagger)}_{\bm{k} \sigma}, f^{(\dagger)}_{\bm{k}, \sigma}$ are annihilation (creation) operators of itinerant $c$ electrons and localized $f$ electrons for momentum $\bm{k}$ and spin $\sigma$, $n_{f j \sigma}=f^{\dagger}_{j \sigma} f_{j \sigma}$ is the number operator for spin $\sigma$ on the $j$-th site, and $\bm{\sigma}=(\sigma_x, \sigma_y, \sigma_z)$ are the Pauli matrices. We consider a square lattice, where the $c$ electrons have a dispersion as $\varepsilon_{\bm{k} \sigma} = -2t(\cos k_x + \cos k_y) + \varepsilon_{c}$, and the $f$ electron band is flat. Here, $t$ is the hopping constant, $\varepsilon_{c (f)}$ is the onsite potential of the $c$($f$) electrons, $U$ is a local interaction between $f$ electrons, $V$ describes a local hybridization between the $c$ and $f$ electrons, and $V'$ describes a nonlocal hybridization between the $c$ and $f$ electrons, which represents a Rashba-type SOC.

We analyze the linear and nonlinear conductivity of this model by using two different methods, namely, the semiclassical Boltzmann equation and the dynamical mean-field theory (DMFT), using Green's functions to calculate the conductivity. The Boltzmann equation gives an intuitive understanding of the scattering of the electrons and the conditions/constraints to observe a nonreciprocal current. However, the Boltzmann equation includes some severe approximations for the interactions and lifetime of the electrons, which lead us to use DMFT to calculate a self-consistent self-energy of this correlated system and the Green's function method to calculate the linear and second-order nonlinear response. 

\section{Boltzmann equation}  \label{Boltzmann}
\subsection{Formalism}

To study the transport properties of this model using the semiclassical Boltzmann equation, we first map the periodic Anderson model on a Kondo lattice model. 
By assuming a half-filled $f$ electron level and $U \gg V, V'$, we describe the $f$ electrons as localized spins. By using second-order perturbation theory\cite{Schrieffer1966} in the local and non-local hybridizations $V$, $V'$, we find the following effective Hamiltonian in leading order:

\begin{eqnarray}
&&H_{\rm{eff}} =H_0 + H_1 + H_2 \\
&&H_0 = \sum_{\bm{k} \sigma} \varepsilon_{\bm{k}} c^{\dagger}_{\bm{k} \sigma} c_{\bm{k} \sigma} \\
&&H_1 = J_1 \sum_{\bm{k} \bm{k'} j} e^{-i(\bm{k'} - \bm{k}) \cdot \bm{r}_j} \bm{S}^c_{\bm{k'} \bm{k}} \cdot \bm{S}^f_j  \label{kondo} \\
&&H_2 = J_2 \sum_{\bm{k} \bm{k'} j} e^{-i(\bm{k'} - \bm{k}) \cdot \bm{r}_j } \bm{g}_{\bm{k'} \bm{k}}\cdot (\bm{S}^c_{\bm{k'\bm{k}}} \times \bm{S}^f_j ) \label{dm} \\
&&\bm{g}_{\bm{k'} \bm{k}} = \bm{d}_{\bm{k'}} - \bm{d}_{\bm{k}}, 
\end{eqnarray}
where $\bm{S}^c_{\bm{k'} \bm{k}} = \frac{1}{2} c^{\dagger}_{\bm{k'} \sigma'} \bm{\sigma}_{\sigma' \sigma} c_{\bm{k} \sigma}$ and $\bm{S}^f_j = \frac{1}{2} f^{\dagger}_{\sigma' j} \bm{\sigma}_{\sigma' \sigma} f_{\sigma j}$ are spin operators, and $J_1=8V^2/U$, $J_2=8VV'/U$.  In addition to a conventional Kondo interaction in Eq.~(\ref{kondo}), we obtain a Dzyaloshinskii-Moriya(DM) interaction between the $c$ and $f$ electrons in Eq.~(\ref{dm}) induced by the Rashba-type SOC. We note that the DM interaction is parity-odd ($\bm{g}_{-\bm{k'} -\bm{k}} = - \bm{g}_{\bm{k'} \bm{k}}$), which can cause a nonreciprocal current.

 We now use the semiclassical Boltzmann equation \cite{Isobe2020,Ishizuka2020} to calculate the conductivity in the Kondo lattice model. In the steady state, the semiclassical Boltzmann equation reads
\begin{eqnarray}
e \bm{E} \cdot \nabla_{\bm{k}} f_{\bm{k} \sigma} &= \sum_{\bm{k'} \sigma'}& \Bigl(W_{\bm{k} \sigma, \bm{k'} \sigma'} f_{\bm{k'} \sigma'} (1 - f_{\bm{k} \sigma}) \nonumber \\ &&- W_{\bm{k'} \sigma', \bm{k} \sigma} f_{\bm{k} \sigma} (1 - f_{\bm{k'} \sigma'}  ) \Bigr),
\end{eqnarray}
where $e<0$ is the elementary charge, and $\bm{E}=(E_x,E_y,E_z)$ is an electrical field. $f_{\bm{{k} \sigma}}$ is the nonequilibrium distribution function which is changed from the equilibrium Fermi distribution function by the electric field and the interactions. We use the first Born approximation on the scattering rate, $W_{\bm{k} \sigma, \bm{k'} \sigma'} = 2 \pi |\bra{\bm{k} \sigma} H_{\mathrm{int}} \ket{\bm{k'} \sigma'}|^2 \delta(\varepsilon_{\bm{k} \sigma} - \varepsilon_{\bm{k'} \sigma'})$, where $c$ electrons are scattered 
by the interaction, $H_{\mathrm{int}}$, from $\ket{\bm{k} \sigma}$ to $\ket{\bm{k'} \sigma'}$. We divide the scattering rate into a symmetric part $W^{+}_{\bm{k} \sigma, \bm{k'} \sigma'}$ and an antisymmetric part $W^{-}_{\bm{k} \sigma, \bm{k'} \sigma'}$ with respect to inversion ($\bm{k} \to -\bm{k}$). Then, we approximate the symmetric part by a relaxation time $\tau$. Previous works \cite{Isobe2020,Ishizuka2020} have shown that the antisymmetric part of the scattering rate can result in a nonreciprocal current. The Boltzmann equation becomes
\begin{equation}
e \bm{E} \cdot \nabla_{\bm{k}} f_{\bm{k} \sigma} = - \frac{f_{\bm{k} \sigma} - f^0_{\bm{k} \sigma}}{\tau}  + \sum_{\bm{k'} \sigma'} W^-_{\bm{k} \sigma, \bm{k'} \sigma'} (f_{\bm{k'} \sigma'} - f_{\bm{k} \sigma}) .
\end{equation}
Here, $f^0_{\bm{k} \sigma} = 1/(e^{\beta \varepsilon_{\bm{k} \sigma}} + 1)$ is the equilibrium Fermi distribution function. 

\subsection{Results} \label{Bolreslut}
We calculate the antisymmetric scattering rate $W^{-}_{\bm{k} \sigma, \bm{k'} \sigma'} = \frac{1}{2} (W_{\bm{k} \sigma, \bm{k'} \sigma'} - W_{-\bm{k} \sigma, -\bm{k'} \sigma'})$ using $H_{\mathrm{int}} = H_1 +H_2$, see Eq. (\ref{kondo}) and (\ref{dm}). The antisymmetric scattering rate reads
\begin{widetext}
\begin{eqnarray} \label{asym}
W^{-}_{\bm{k} \sigma, \bm{k'} \sigma'} 
&=& -2 \pi i J_1 J_2 \left\{ \bm{g}_{\bm{k'} \bm{k}} \cdot \bm{M}  - \sigma \delta_{\sigma \bar{\sigma'}} \sum_{j l}\cos((\bm{k'}-\bm{k}) \cdot (\bm{r}_j - \bm{r}_l)) (g_x \{ S^f_{zl}, S^f_{xj} \} - g_y \{ S^f_{zl}, S^f_{yj} \}) \right\} \delta(\varepsilon_{\bm{k}} - \varepsilon_{\bm{k'}})  . \nonumber \\
\end{eqnarray}
\end{widetext}

Here, $\bm{M} = \sum_j \langle\bm{S}^f_j\rangle$ is the  magnetization of the localized $f$ electrons,  $\bm{g}_{\bm{k'k}}=(g_x,g_y,g_z)$, $\{A,B\}=AB+BA$ the anti-commutator, and $\bar{\sigma} = - \sigma$. The coefficient $J_1J_2$ in Eq.~(\ref{asym}) demonstrates that the conventional Kondo interaction and the DM interaction are necessary to generate a finite antisymmetric scattering rate. 

Because the second term is proportional to the magnetization of the noninteracting $c$ electrons, $\sim\sigma$, it can be expected to be small compared to the first term involving the magnetization of the strongly correlated $f$ electrons. Thus, in the following, we focus on the first term in Eq. (\ref{asym}). This assumption is later confirmed using DMFT. We then solve the Boltzmann equation and calculate the nonequilibrium  distribution function by expanding in the order of the electric field and the antisymmetric scattering rate;
$f_{\bm{k} \sigma} = f^0_{\bm{k} \sigma} + \sum_{ij} f^{(i,j)}_{\bm{k} \sigma}$, where $f^{(i,j)}_{\bm{k} \sigma}$ includes the $i$-th order of $E$ and the $j$-th order of $W^-$. The electric current  calculated in the first order of $\bm{M}$ is then given by
\begin{equation}
\bm{J} = e \sum_{\bm{k} \sigma} \frac{\partial \varepsilon_{\bm{k}}}{\partial \bm{k}} f_{\bm{k} \sigma}.
\end{equation}
For simplicity, we assume that the $c$ electron band has a hyperbolic dispersion, $\varepsilon_{\bm{ik}} \simeq \frac{k^2}{2m} -\mu$ ($\mu>0$), and the Rashba-type SOC is linearized, $\bm{g}_{\bm{k'} \bm{k}} \sim -i (k'_y - k_y , -k'_x + k_x)$. This assumption is justified when the $c$ electrons are low-filled.  We later confirm the existence of a ferromagnetic phase for this situation using DMFT.

A symmetry analysis of the system within the Boltzmann equation yields several restrictions for the conductivity. When $M_x \neq 0, M_y = M_z =0$, the system has mirror symmetry, $(x,y,z) \to (-x,y,z)$. Thus, the conductivity containing an odd number of $x$-indices must vanish. The same holds for the $y-$direction. In addition, the Hamiltonian has another mirror symmetry, $\mathcal{M} : (x,y,z) \to (-x,-y,z)$, 
$\mathcal{M} H(M_x,M_y,M_z) \mathcal{M}^{-1} = H(-M_x,-M_y,M_z)$. This symmetry restricts the dependence of the conductivity on the magnetization as $\sigma_{ij} (M_x,M_y,M_z) = \sigma_{ij} (-M_x,-M_y,M_z)$ and $\sigma_{ijk} (M_x,M_y,M_z) = - \sigma_{ijk} (-M_x,-M_y,M_z)$ , where $i,j,k\in\{x, y\}$. Thus, the linear conductivity is even in $M_x$ and $M_y$, and the second-order conductivity is odd. The linear conductivity is hardly affected by the magnetization because it depends in leading-order on $\bm{M}^2$. Furthermore, the second-order conductivity can only be nonzero in the presence of a ferromagnetic magnetization.

Calculating the conductivity using the Boltzmann equation, we use that the current is related to the conductivity by $J_{i} = \sigma_{ij} E_j + \sigma_{ijk} E_j E_k + \cdots$. Then, the linear conductivity becomes
\begin{equation}
\sigma^{(1)} = 
\left(
    \begin{array}{cc}
      \sigma_{xx} & \sigma_{xy} \\
      \sigma_{yx} & \sigma_{yy}
    \end{array}
  \right)
\simeq
\left(
    \begin{array}{cc}
      \frac{\tau e^2 \mu}{ \pi} & 0 \\
      0 & \frac{\tau e^2 \mu}{ \pi}
    \end{array}
  \right),
\end{equation}
and the second-order conductivity becomes
\begin{eqnarray}
\sigma^{(2)} &=& 
\left(
    \begin{array}{cc}
      \sigma_{xxx} & \sigma_{xyy} \\
      \sigma_{yxx} & \sigma_{yyy}
    \end{array}
  \right) \nonumber \\
&\simeq&
\frac{6 m \mu}{\pi} (\tau e)^3 J_1 J_2  
\left(
    \begin{array}{cc}
      M_y & M_y \\
      -M_x & -M_x
    \end{array}
  \right) .
\end{eqnarray}
 
We see that the nonlinear conductivity depends on the product of $J_1=8V^2/U$ and $J_2=8VV^\prime/U$. 
An interesting feature of the second-order conductivity is that it becomes only finite in the direction perpendicular to the magnetization without regard for the direction of the electric field.  Thus, we can control the direction of the nonreciprocal current by controlling the direction of the magnetization.
This characteristic is unique to a bulk system, which cannot be realized in a junction.

\section{Microscopic analysis using DMFT}  \label{Micro}
\subsection{DMFT and the Green's function method for the conductivity}
Because the Boltzmann equation includes some severe approximations for the interactions and lifetime of the electrons, we next use DMFT \cite{Metzner1989,Georges1996,Pruschke1995} to calculate a self-consistent self-energy of this correlated system and the Green's function method to calculate the linear and second-order nonlinear conductivity\cite{Michishita2021}. 
DMFT maps the lattice model onto a quantum impurity model that has to be solved self-consistently.
To calculate the self-energy of the impurity model, we use the numerical renormalization group (NRG) \cite{Wilson1975,Bulla2008}. NRG is an accurate tool to calculate self-energies of correlated impurity models with high accuracy at the Fermi energy\cite{Peters2006,Weichselbaum2007}.
DMFT combined with the Green's function method is a valuable tool for microscopic analysis of correlated electron systems.

After having calculated a self-consistent self-energy of the system in the ferromagnetic state, we analyze the linear and nonlinear conductivity.
We calculate the conductivity using the Green's function method in both the ordered phase and disordered phase. At finite temperature, the linear DC conductivity and the second-order DC conductivity read \cite{Michishita2021}
\begin{widetext}
\begin{eqnarray}
\sigma_{ij} &=& \int^{\infty}_{-\infty} \frac{d \omega}{2 \pi} \int_{\mathrm{BZ}} \frac{d \bm{k}}{(2 \pi)^2} \left\{ \left( - \frac{\partial f(\omega)}{\partial \omega} \right) \mathrm{Re} \mathrm{Tr} \left( J_i  G^R(\omega) J_j G^A(\omega) \right) -2 f(\omega) \mathrm{Re} \mathrm{Tr} \left( J_i  \frac{\partial G^R(\omega)}{\partial \omega} J_j G^R(\omega) \right) \right\}  \nonumber \\\label{EQlinear}
\\
\sigma_{ijk} &=& -4 \int^{\infty}_{-\infty} \frac{d \omega}{2 \pi} \int_{\mathrm{BZ}} \frac{d \bm{k}}{(2 \pi)^2} \left\{ \left( - \frac{\partial f(\omega)}{\partial \omega} \right) \mathrm{Im} \mathrm{Tr} \left( J_i  \frac{\partial G^R(\omega)}{\partial \omega} J_j G^R(\omega) J_k G^A(\omega) +J_i  \frac{\partial G^R(\omega)}{\partial \omega} J_{jk} G^A(\omega) \right)  \right\}  \nonumber \\
&& +2 \int^{\infty}_{-\infty} \frac{d \omega}{2 \pi} \int_{\mathrm{BZ}} \frac{d \bm{k}}{(2 \pi)^2} \left\{ f(\omega) \mathrm{Im} \mathrm{Tr} \left( 2 J_i  \frac{\partial}{\partial \omega} \left( \frac{\partial G^R(\omega)}{\partial \omega} J_j G^R(\omega) \right) J_k G^R(\omega)  + J_i \frac{\partial^2 G^R(\omega)}{\partial \omega^2} J_{jk} G^R(\omega) \right) \right\}, \nonumber \\\label{EQnonlinear}
\end{eqnarray}
\end{widetext}
where $f(\omega) = 1/(e^{\beta \omega} + 1)$ is the Fermi distribution function, $J_i = \partial H(\bm{k})/ \partial k_i$ and $J_{ij} = \partial^2 H(\bm{k}) / \partial k_i \partial k_j$ are velocity operators, and $G^{R/A}$ are the retarded and advanced Green's functions. 
This equation includes correlation effects via the self-energy, $G^{R}(\omega, \bm{k}) = 1/(\omega - H_0(\bm{k}) - \Sigma(\omega))$. However, it neglects the momentum dependence of the self-energy and vertex corrections. Furthermore, we add a small imaginary constant $i \eta=i 10^{-3}$ to the self-energy, which can be interpreted as a dissipation effect arising from impurities. 

\subsection{Ferromagnetism}
\begin{figure*}[t]
\includegraphics[width=0.24\linewidth]{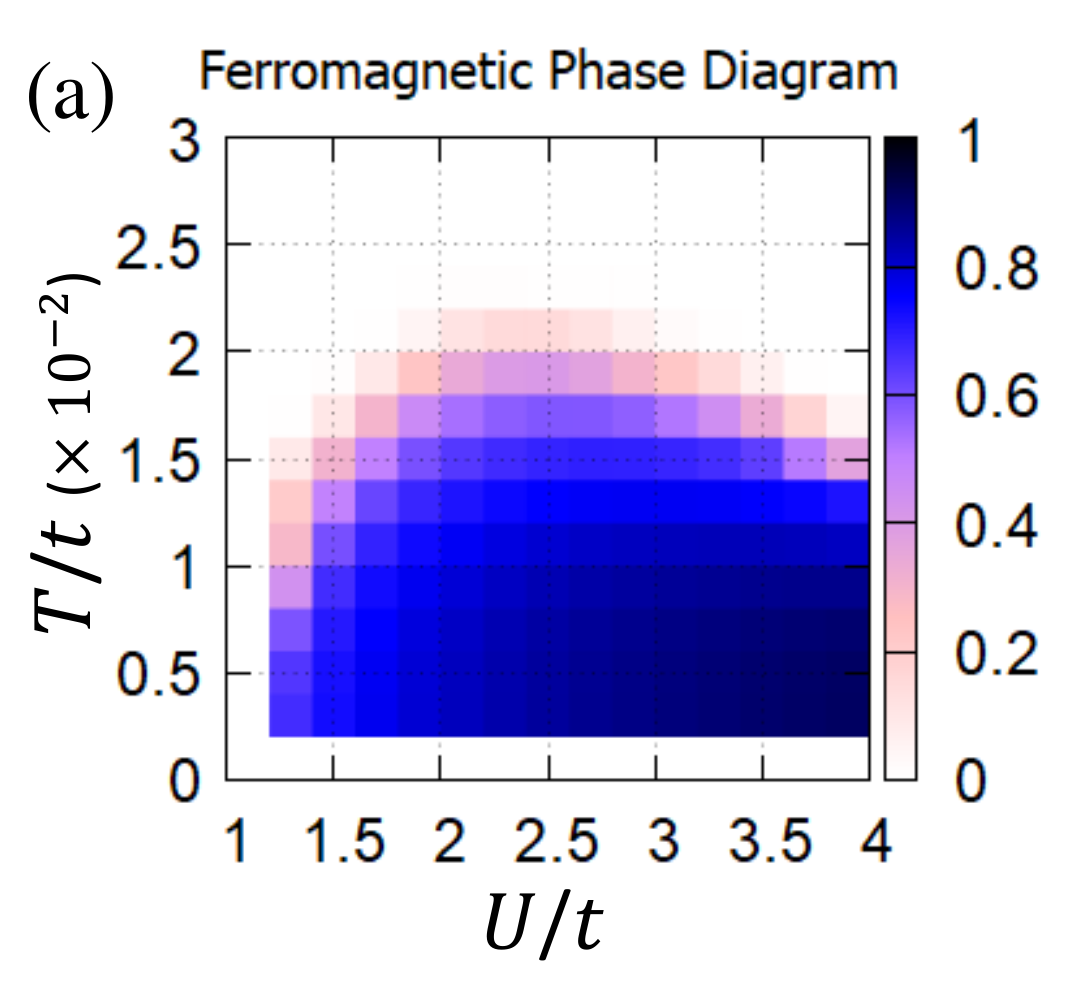}
\includegraphics[width=0.24\linewidth]{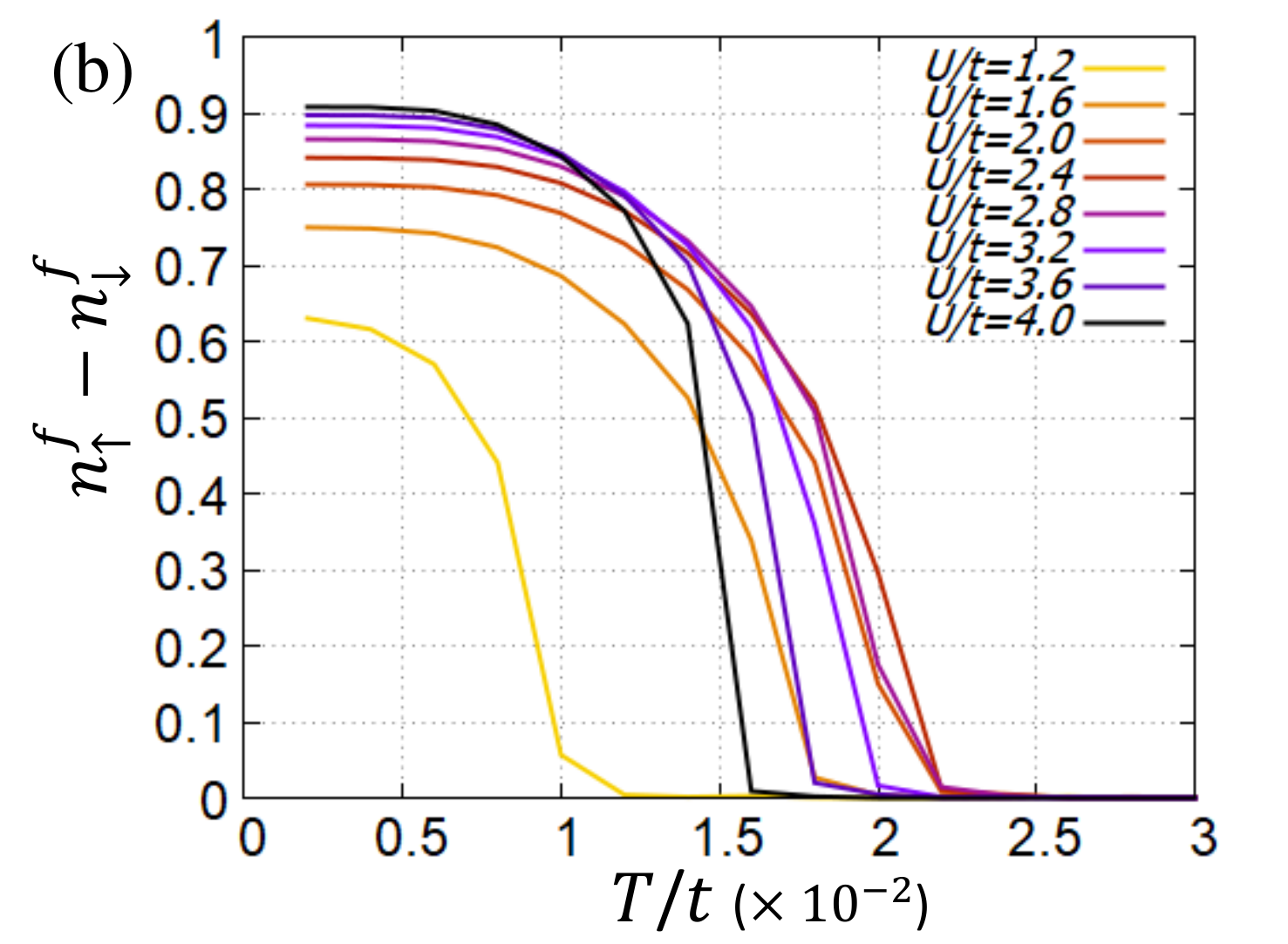}
\includegraphics[width=0.24\linewidth]{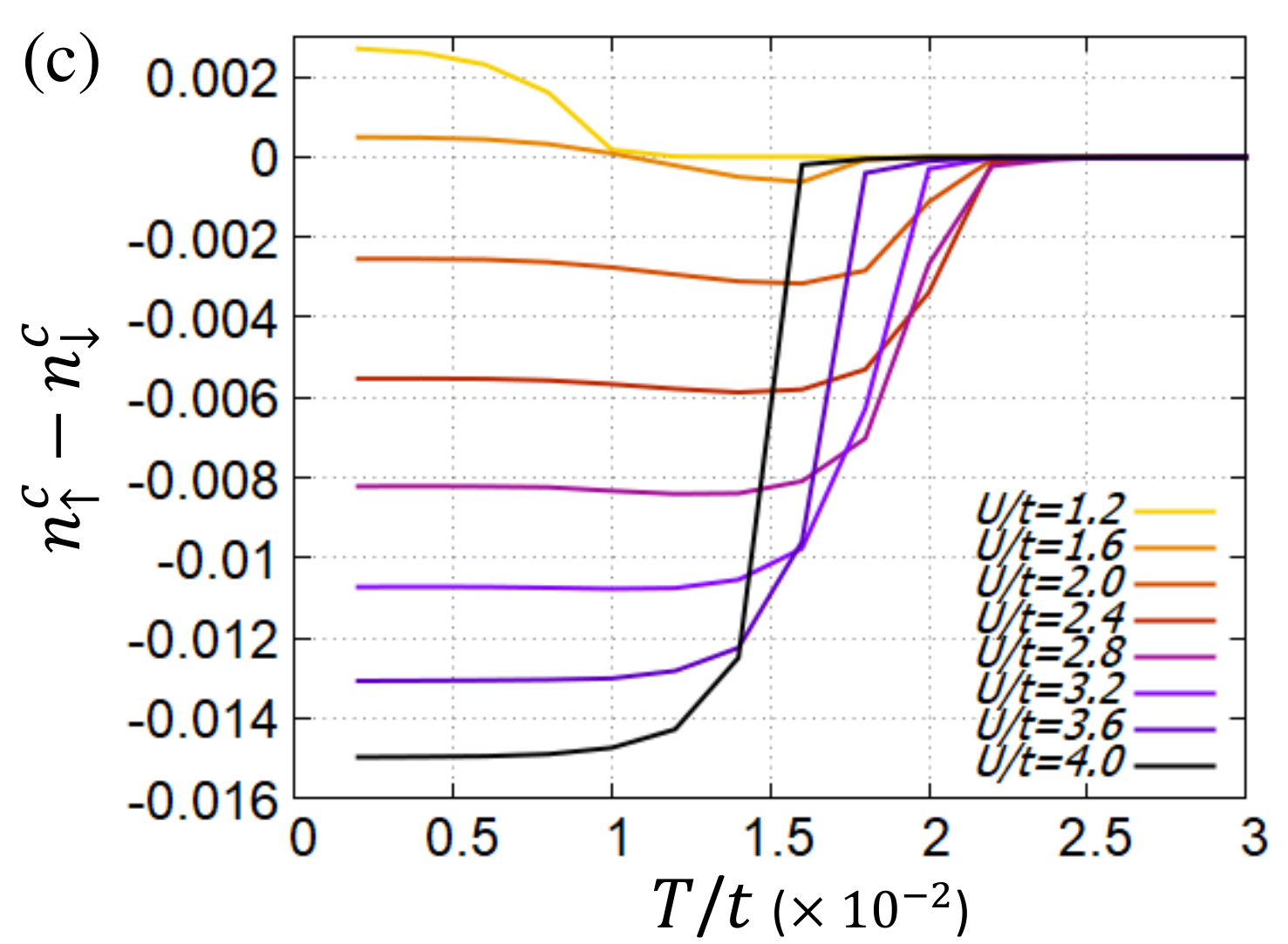}
\includegraphics[width=0.24\linewidth]{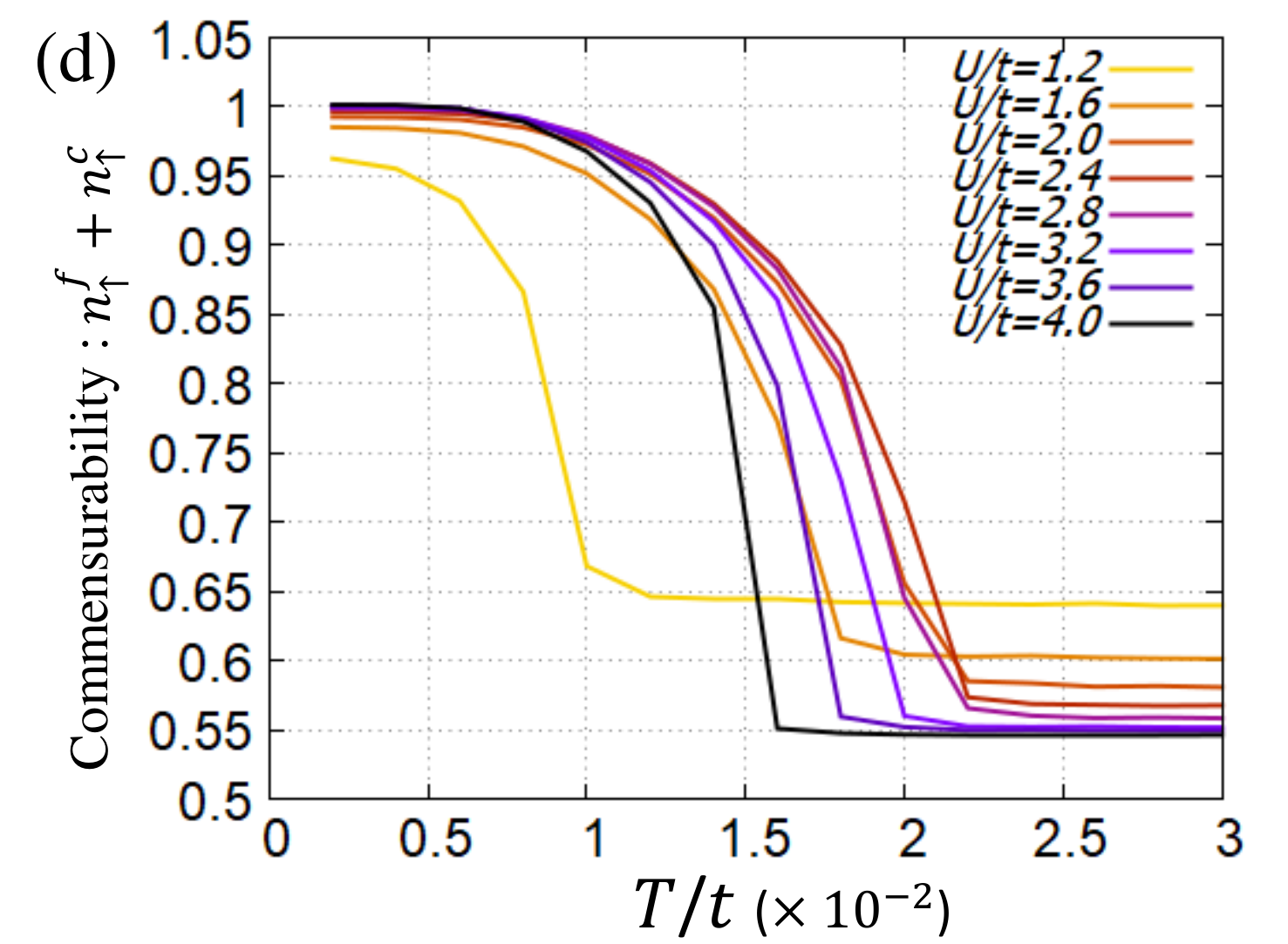}
\caption{\label{mag} (a) Ferromagnetic phase diagram showing the magnetization of the $f$ electrons for various temperatures, $T$, and interaction strengths, $U$. The direction of the magnetization is in the $x$-direction.  Panels (b) and (c) show  the magnetization in the $x$-direction of the $f$ electrons, $n^f_{\up}-n^f_{\down}$, and $c$ electrons $n^c_{\up}-n^c_{\down}$, respectively. In panel (d), we show  $n^c_{\up}+n^f_{\up}$, which is called commensurability in the main text.  The data is obtained using DMFT for $\varepsilon_c/t=3.6$, $\varepsilon_f=-U/2$, $V/t=1.0$, $V'/t=0.1$ and $t=0.05$.}
\end{figure*}
\begin{figure}[t]
\centering
\includegraphics[width=\linewidth]{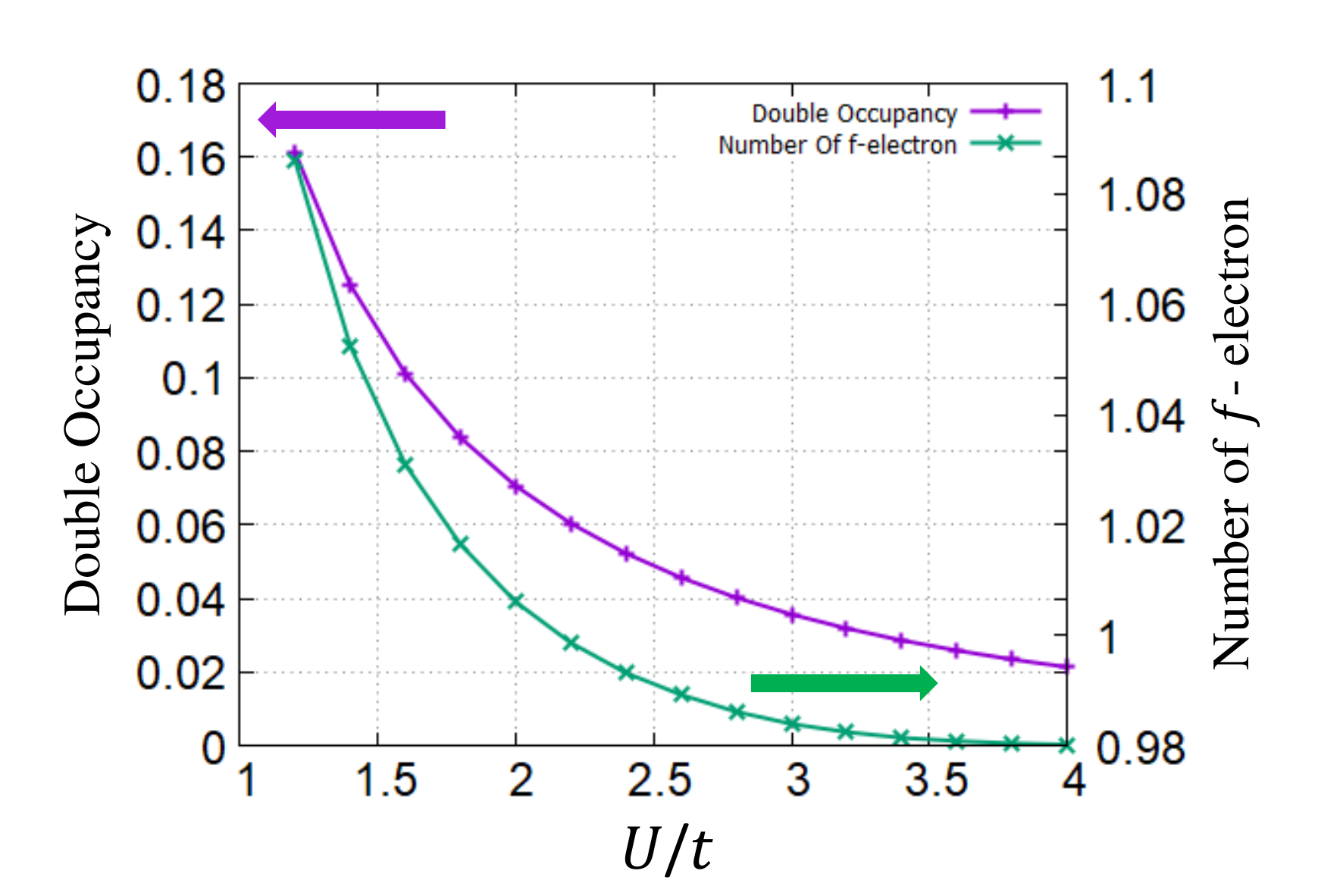}
\caption{\label{double}  Double occupancy of the $f$ electrons $\langle n^f_{\up} n^f_{\down} \rangle$ (left axis) and the number of $f$ electrons $n^f_{\up}+n^f_{\down}$ (right axis) at $T/t=0.2 \times 10^{-2}$. Parameters are the same as in Fig.~\ref{mag}.}
\end{figure}
\begin{figure*}[t]
\includegraphics[width=0.24\linewidth]{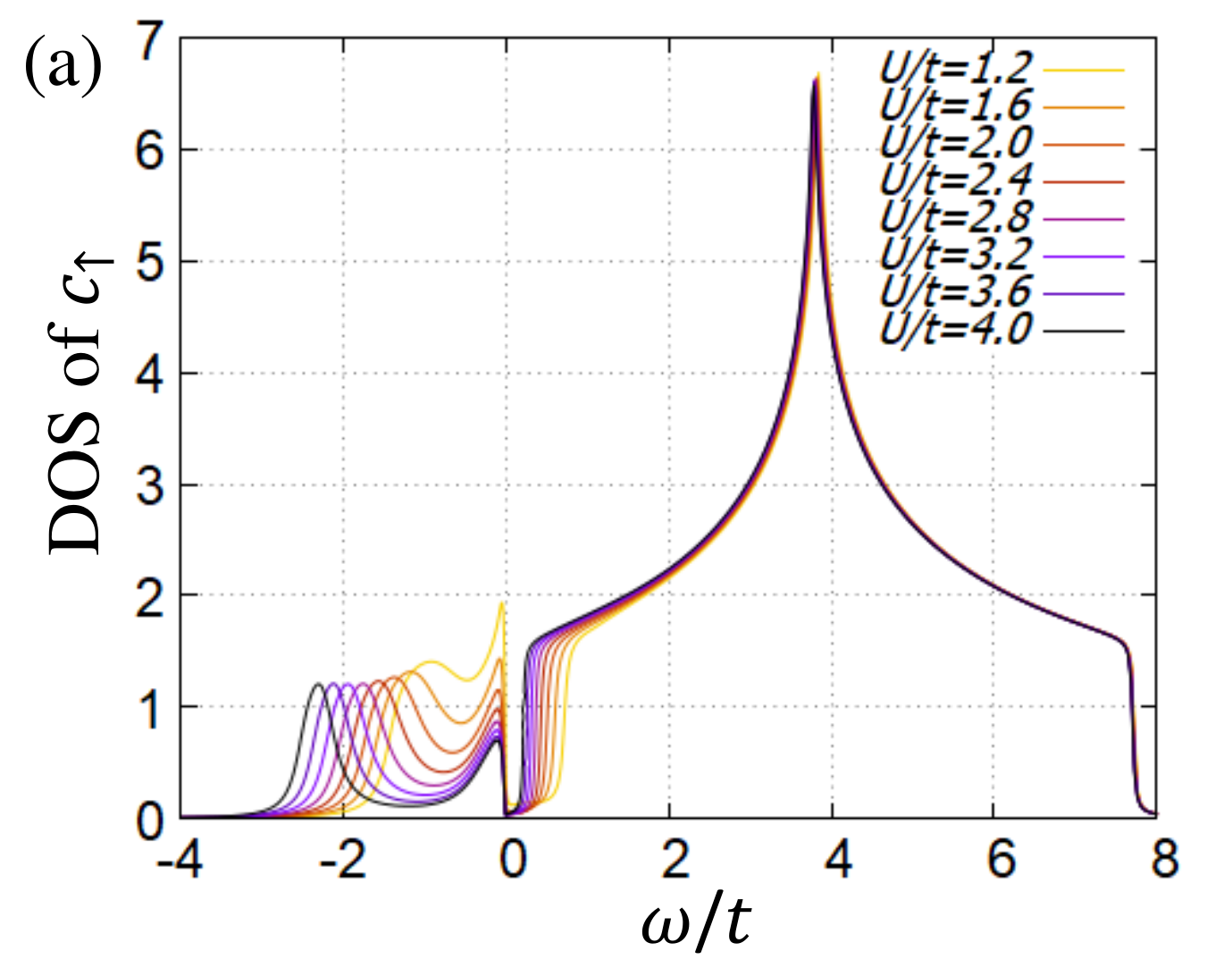}
\includegraphics[width=0.24\linewidth]{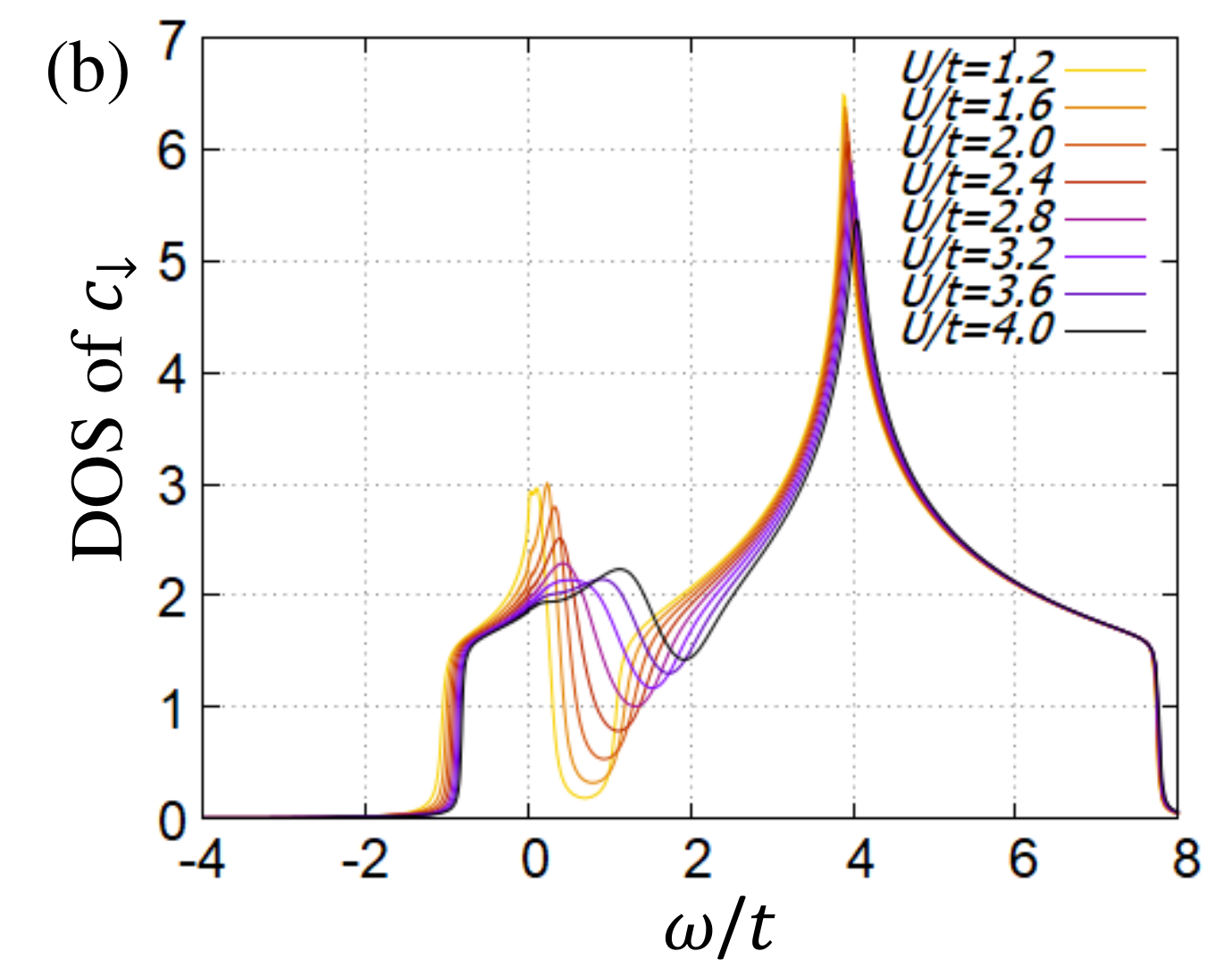}
\includegraphics[width=0.24\linewidth]{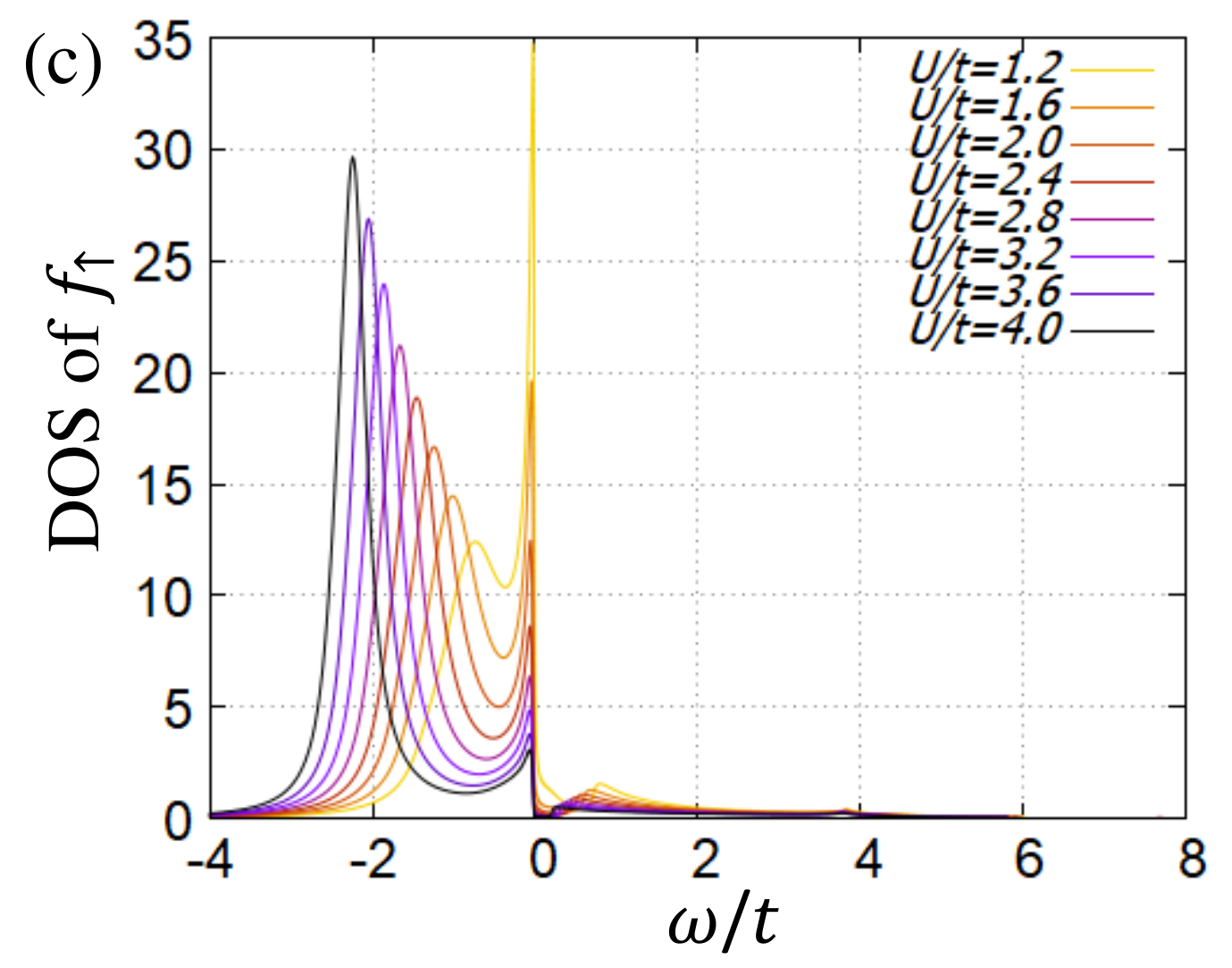}
\includegraphics[width=0.24\linewidth]{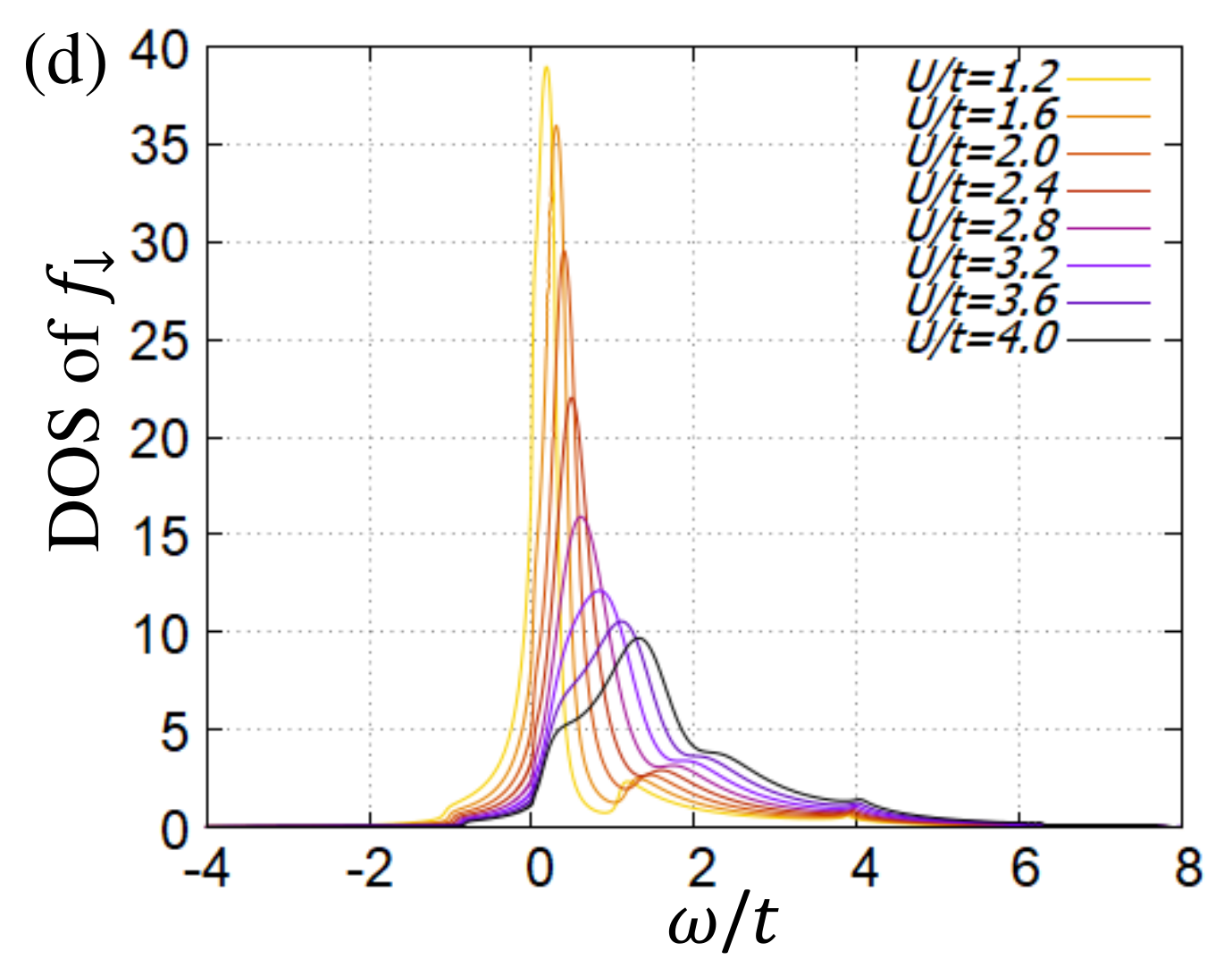}
\caption{\label{dos} Density of states (DOS) of the (a) $c_{\up}$-electrons, (b) $c_{\down}$-electrons, (c) $f_{\up}$-electrons and (d) $f_{\down}$-electrons. Parameters are the same as in Fig.~\ref{mag}. The temperature is $T/t=0.2 \times 10^{-2}$. }
\end{figure*} 
\begin{figure*}[t]
\includegraphics[width=0.325\linewidth]{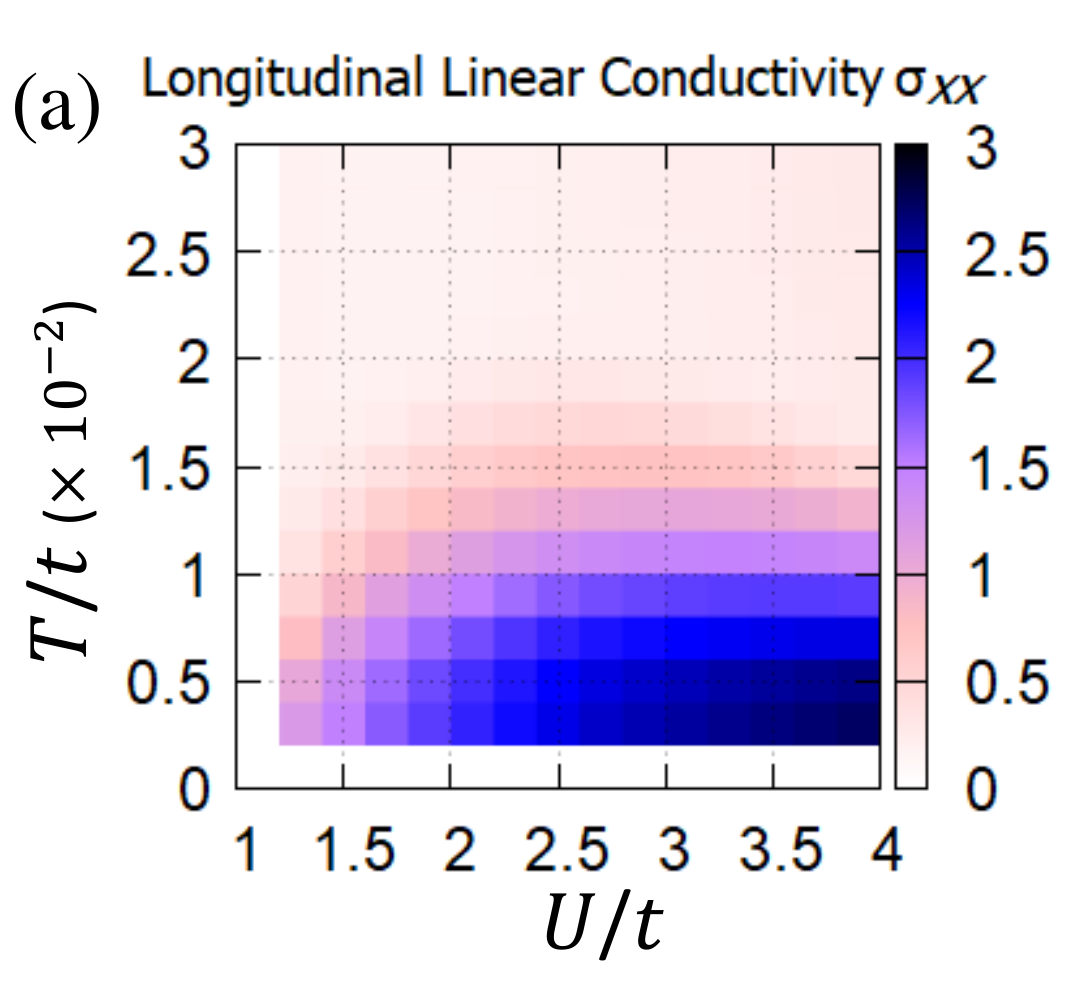}
\includegraphics[width=0.325\linewidth]{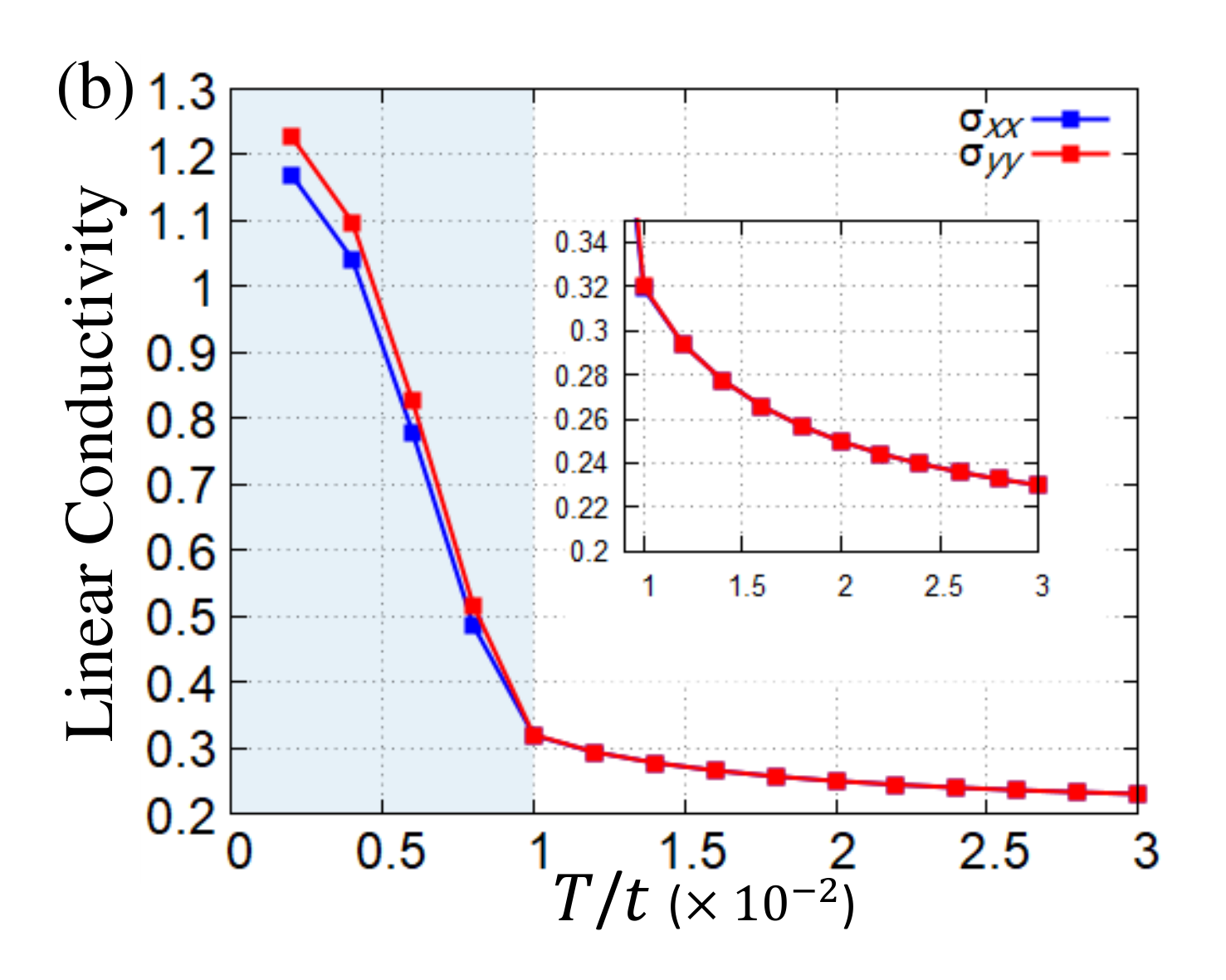}
\includegraphics[width=0.325\linewidth]{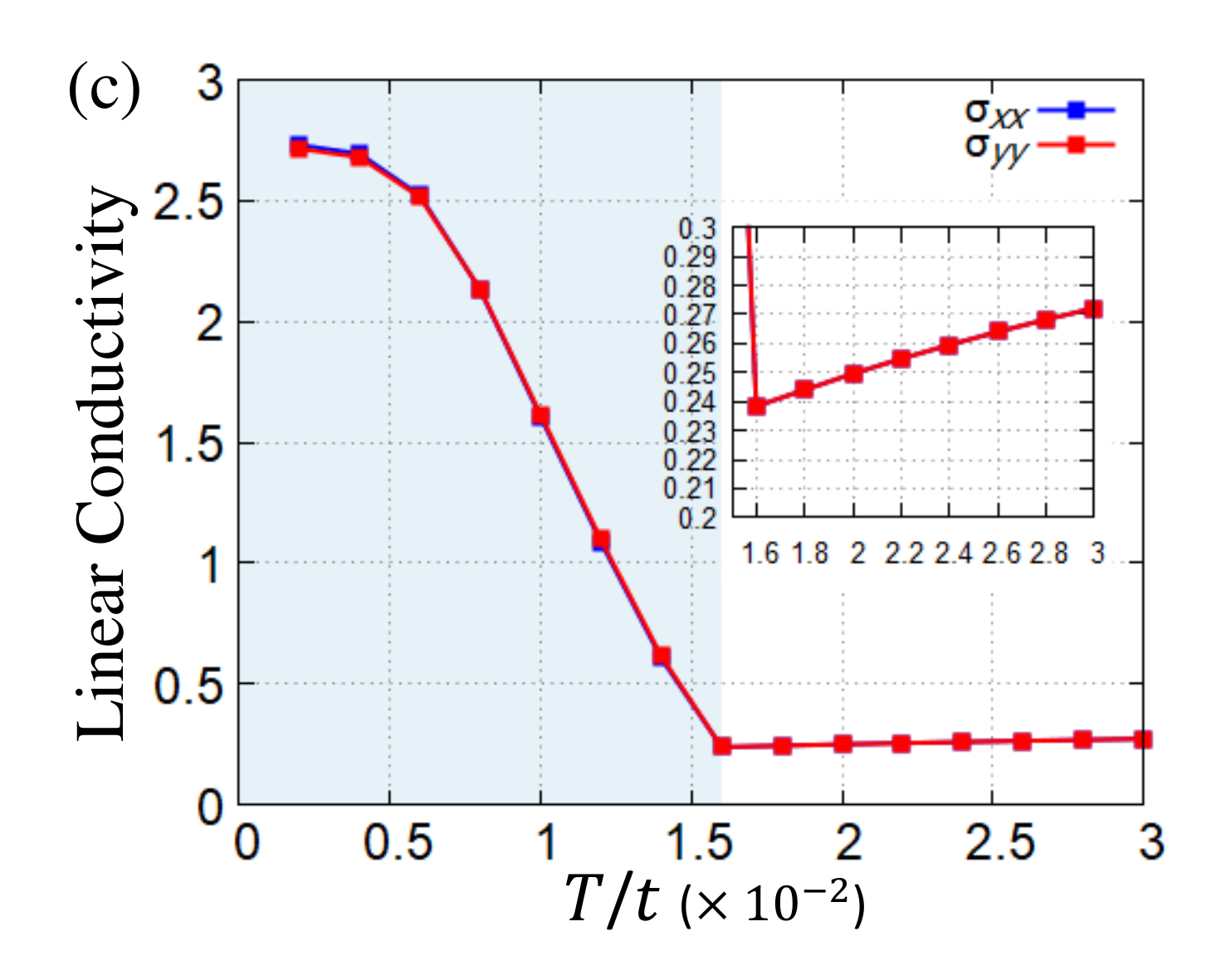}
\caption{\label{linear}(a) Longitudinal linear conductivity, $\sigma_{xx}$, for various temperatures and interaction strengths. For a better visibility, we show the conductivities $\sigma_{xx}$ and $\sigma_{yy}$ for $U/t=1.2$ (b) and  $U/t=4.0$ (c).
The shaded region corresponds to the ferromagnetic phase. The insets are magnifications above the ferromagnetic transition temperature. Parameters are the same as in Fig.~\ref{mag}.  }
\end{figure*}

Motivated by the results obtained from the semiclassical Boltzmann equation, we study the ferromagnetic phase in a two-dimensional periodic Anderson model, including a Rashba-type SOC using DMFT. We note that the Rashba-type SOC is included in the single-particle part of the Hamiltonian, affecting various physical quantities.
We use the following parameters: $\varepsilon_c/t=3.6$, $\varepsilon_f=-U/2$, $V/t=1.0$, $V'/t=0.1$, and vary the interaction strength, $U$, and the temperature, $T$.

In Fig.~\ref{mag}(a), we show the ferromagnetic phase diagram with magnetization in the $x$-direction of the $f$ electrons
for various temperatures and interaction strengths. The number of $c$ electrons for these parameters smoothly changes between $0.13$ ($U/t$=4.0) and $0.21$ ($U/t$=1.2).
The ferromagnetic transition temperature takes the maximum value around $U/t=2.4$. The transition temperature decreases from $U/t=2.4$ with decreasing interaction strength, and the order parameter vanishes at $U/t=1$. At weak interaction strengths, the double occupancy of the $f$ electrons, $\langle n^f_{\up} n^f_{\down} \rangle$, is large, as shown in Fig.~\ref{double}. Thus, the $f$ electrons strongly hybridize with the $c$ electrons and are itinerant. This absence of local magnetic moments explains the vanishing of the ferromagnetic phase at weak interactions. 
On the other hand,  increasing the interaction strength from $U/t=2.4$, we see that the transition temperature also slowly decreases. In this region, the double occupancy is small, and the $f$ electrons are half-filled. Thus, in this regime, $f$ electrons are localized and can be well described as localized spins as assumed in the Kondo lattice model. However, as the exchange interaction between  $c$ and $f$ electrons is proportional to $1/U$, the transition temperature decreases for strong interaction strengths.
The obtained phase diagram resembles the Doniach phase diagram \cite{Doniach1977}. 

In Figs.~\ref{mag}(b) and (c), we show the magnetization of the $f$ and $c$ electrons in the ferromagnetic phase for different temperatures and interaction strengths.
 We clearly see that the ferromagnetic transition of the $c$ electrons and  $f$ electrons occurs at the same temperature. However, the magnetization of the $f$ electrons is much larger than that of the $c$ electrons, justifying the approximation made for the asymmetric scattering rate in the Boltzmann equation above.
 The magnetization of the $c$ electrons is small because they are noninteracting and low-filled.

Previous studies \cite{Tsunetsugu1997,Wagner2020,PhysRevB.92.094401,Smerat2009,JPSJ.84.094702,Li2010,Liu2013,Peters2012,Peters2017,PhysRevB.88.054431} have shown that a ferromagnetic phase exists in the  SU(2)-symmetric Kondo lattice model for low-filled $c$ electrons at zero temperature. Our results obtained here show that a ferromagnetic phase with magnetization in the $x$-direction exists even for weak Rashba-type SOC in the periodic Anderson model.
However, we note that DMFT does not converge for strong Rashba-type SOC, and the ferromagnetic phase disappears. More complicated magnetic states, such as a chiral magnet, can be expected in the strong SOC regime. However, the analysis of magnetic states in the strong SOC regime goes beyond the scope of this study.

In Fig.~\ref{dos}, we show the density of states (DOS) of the $c$  and $f$ electrons with spin quantization axis in the $x$-direction in the ferromagnetic phase  at $T/t=0.2 \times 10^{-2}$ for various interaction strengths. Consistent with Fig.~\ref{mag}, we observe that most spectral weight of the $f$ electrons below the Fermi energy is in the $x-$up direction. We see sharp peaks in the $f$ electron spectrum near the Fermi energy, especially at weak interaction strengths.
These peaks are formed below the Kondo temperature by hybridization between the $c$ and $f$ electrons. Due to the localization of the $f$ electrons, the hybridization between the $c$ and $f$ electrons becomes weak for strong interaction strengths, and these peaks diminish. Previous  works\cite{Li2010,Liu2013} have found similar results showing that itinerant $f$ electrons coexist in the ferromagnetic phase of the Kondo lattice model at large Kondo coupling corresponding to weak Coulomb interaction in our calculations.

Furthermore, we observe a small gap at the Fermi energy ($\omega=0$) in the spectrum of the $c$ and $f$ electrons for $x-$up in Fig.~\ref{dos} (a) and (c). 
This is consistent with previous works \cite{Peters2012,PhysRevB.88.054431,PhysRevB.86.165107}, where it was called a spin-selective Kondo insulator.
In fact, Fig.~\ref{mag}(d) shows that the commensurability condition ($n^x_{f\up} + n^x_{c \up} =1$) is fulfilled at low temperature and strong interaction strengths in the ferromagnetic phase. Thus, the spectrum for the electrons with $x$-up spin opens a gap because it is half-filled, and we can expect that physical quantities dominated by the Fermi surface, like the conductivity, will strongly depend on the spin components.

\subsection{Conductivities}
Next, we calculate the longitudinal linear conductivity for various temperatures and interaction strengths. We show the linear conductivity $\sigma_{xx}$ over $U$ and $T$ in Fig.~\ref{linear}(a). For better visibility, we show the dependence of the conductivities $\sigma_{xx}$ and $\sigma_{yy}$ on the temperature for $U/t=1.2$ (b) and $U/t=4.0$ (c). It becomes clear that the linear conductivity is enhanced by ferromagnetism and increases with increasing magnetization. 

This is rather remarkable, as the DOS around the Fermi energy decreases with increasing interaction strength, as shown in Fig.~\ref{dos}. However, this increase in the linear conductivity can be explained by a change of the Fermi surface due to the ferromagnetic magnetization and the Kondo effect. The energetical shift of the bands for different spin directions and the hybridization between $c$ and $f$ electrons results in a Fermi surface increasing the product of the velocity operator and the Green's function at the Fermi energy in Eq.~(\ref{EQlinear}), thus, increasing the conductivity.
Furthermore, we see that $\sigma_{xx}$ and $\sigma_{yy}$ develop a  small difference in the ferromagnetic phase, as shown in Fig.~\ref{linear}(b) and (c), while they are equal in the paramagnetic phase.
This difference can be explained by the magnetization in the $x$-direction and the Rashba SOC, resulting in different scattering processes in the $x$- and $y$-direction.
In Fig.~\ref{linear}(b) (and its inset), which shows the conductivity for small interaction strengths, we see that the linear conductivity 
 increases with decreasing temperature above the transition temperature.
This increase can be explained by a Kondo temperature that is larger than the ferromagnetic transition temperature. Thus, the scattering rate decreases with decreasing temperature, resulting in an increase in the conductivity.
On the other hand, in Fig.~\ref{linear}(c) (and its inset), which shows the conductivity for a strong interaction strength, the conductivity above the transition temperature decreases with decreasing temperature. The Kondo temperature is below the ferromagnetic transition temperature so that Kondo scattering is strong. This difference in the conductivities also agrees with the conclusions drawn above from the density of states.

\begin{figure}
\centering
\includegraphics[width=6.0cm]{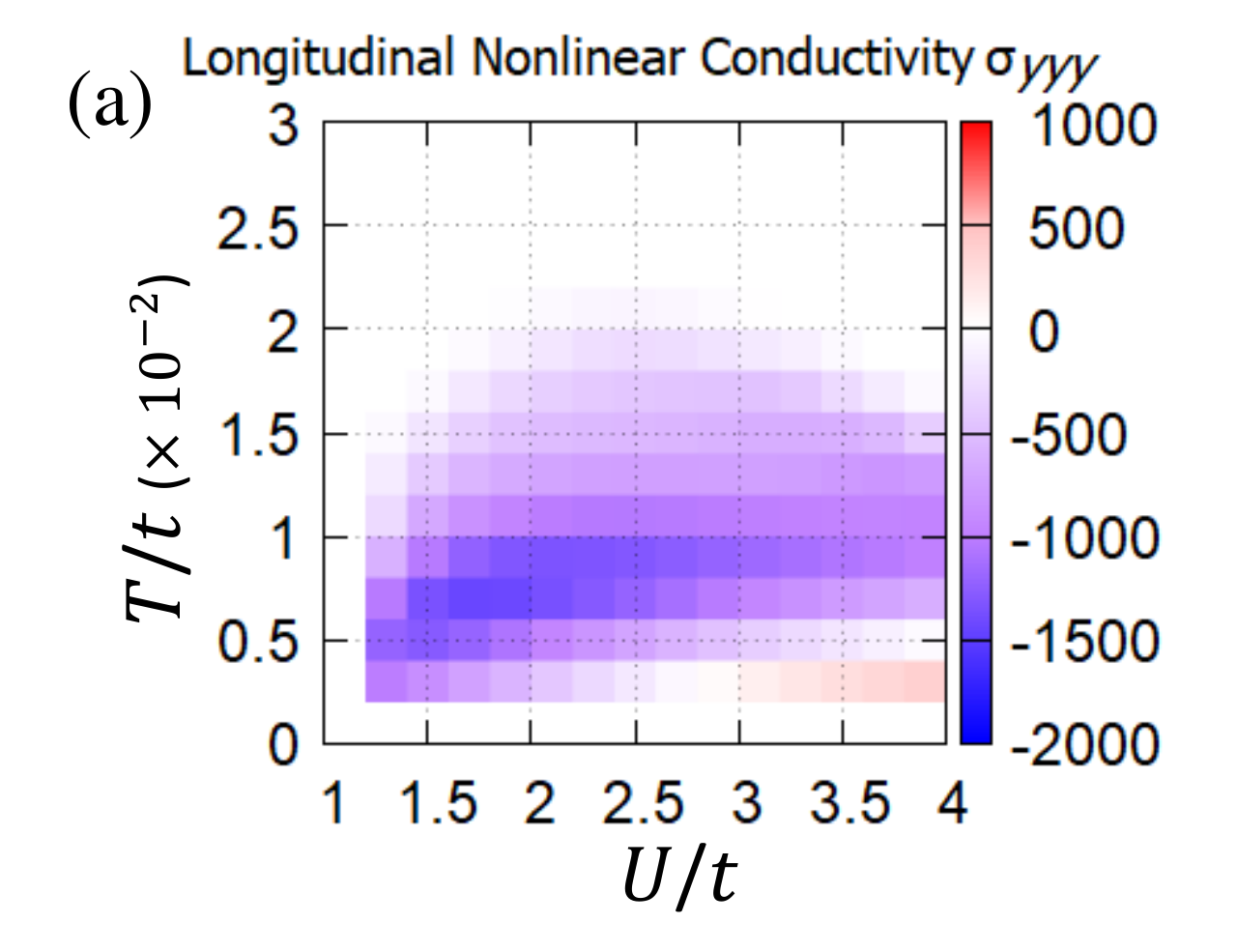}
\includegraphics[width=6.0cm]{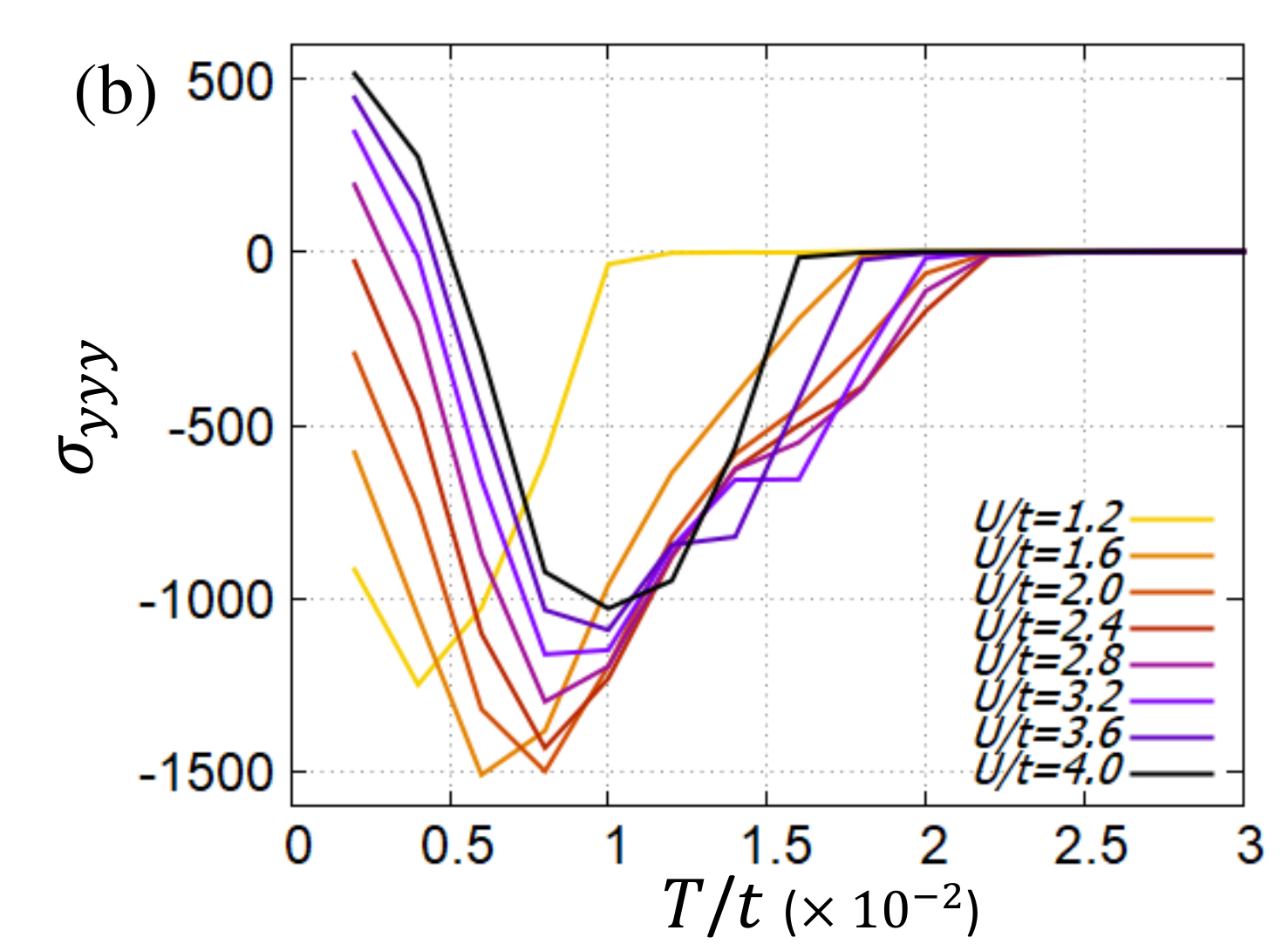}
\caption{\label{nonlinear}  (a) Longitudinal second-order nonlinear conductivity for various temperatures and interaction strengths. For a better visibility, we show the nonlinear conductivity for a set of interaction strengths in (b). Parameters are the same as in Fig.~\ref{mag}.}
\end{figure}

Next, we calculate the longitudinal nonlinear conductivity at various temperatures and interaction strengths, as shown in Fig.~\ref{nonlinear}. It becomes clear that a second-order conductivity appears only in the ferromagnetic phase. Furthermore, for magnetization in the $x$-direction, only the longitudinal nonlinear conductivity $\sigma_{yyy}$ is finite.
This is due to the symmetry constraints, as discussed in Sec.~\ref{Bolreslut}. The second-order nonlinear conductivity can only occur perpendicular to the magnetization. 
We show the details of the nonlinear conductivity depending on the temperature in Fig.~\ref{nonlinear}(b). It is immediately apparent that the nonlinear conductivity first becomes finite at high temperatures in the ferromagnetic phase and is negative. Decreasing the temperature, we see that the magnitude of the nonlinear conductivity reaches a maximum at a finite temperature. When the temperature is further decreased, the magnitude of the nonlinear conductivity decreases. For strong interaction strengths, the sign of the nonlinear conductivity changes at very low temperatures. We will analyze this sign change in more detail below.

\begin{figure}
\centering
\includegraphics[width=7.0cm]{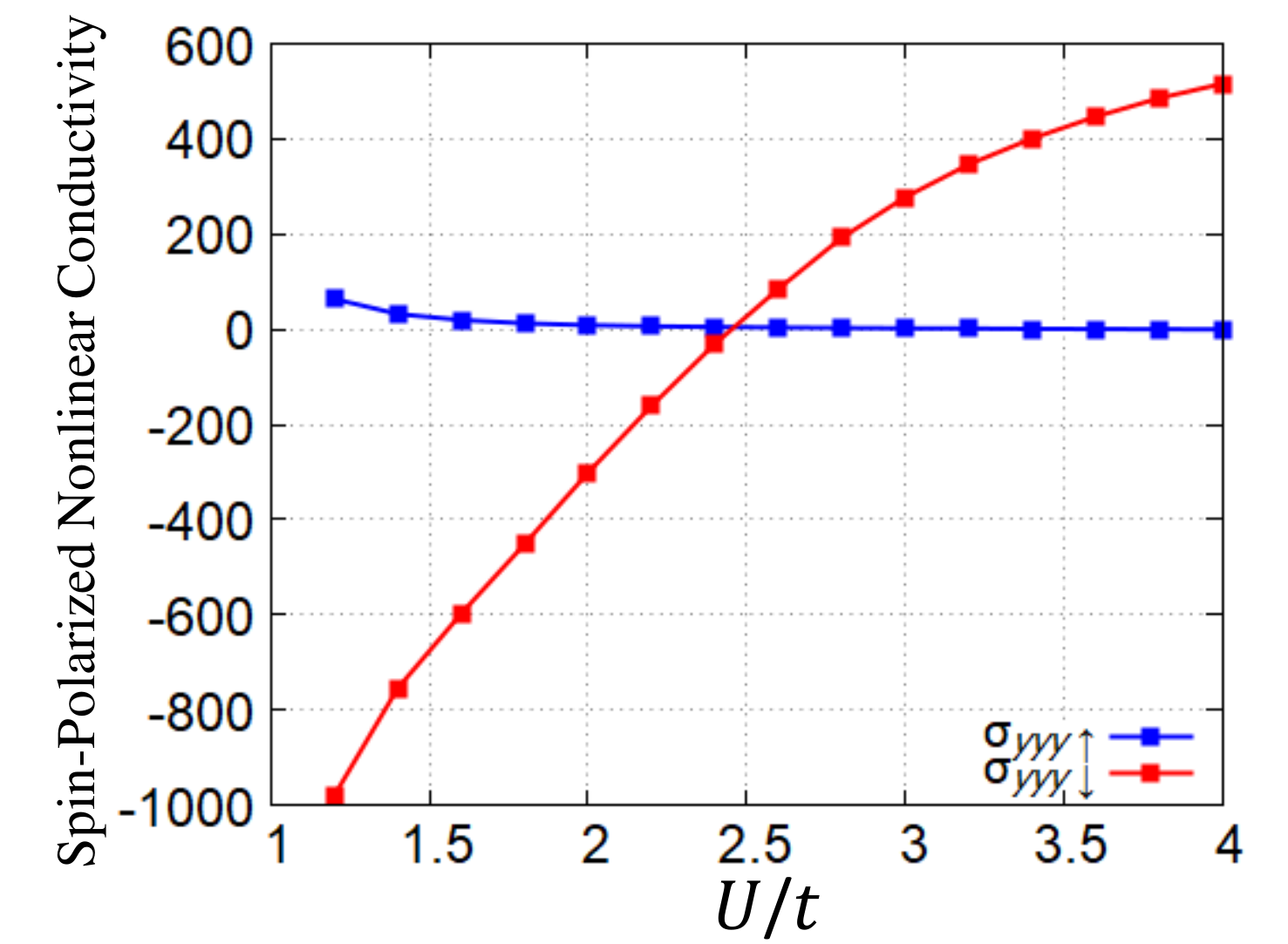}
\caption{\label{spinpolar} 
Nonlinear contribution to the spin-polarized current  $\langle J_{y \up (\down)}\rangle$ at $T/t=0.2 \times 10^{-2}$. Parameters are the same as in Fig.~\ref{mag}.}
\end{figure}

Furthermore, the gap in the DOS of the $x$-up electrons for strong interaction strengths and low temperatures influences the linear and nonlinear conductivity resulting in spin-dependent conductivities.
In order to analyze this spin dependence, we calculate the nonlinear conductivity for spin polarized currents
$J_{y \up} \equiv  \sum_{\bm{k}} ( 2t \sin k_y c_{\bm{k} \up}^{\dagger} c_{\bm{k} \up} +  V' \cos k_y (c_{\bm{k} \up}^{\dagger} f_{\bm{k} \up} + f_{\bm{k} \up}^{\dagger} c_{\bm{k} \up}) )$ and $J_{y \down} \equiv  \sum_{\bm{k}} ( 2t \sin k_y c_{\bm{k} \down}^{\dagger} c_{\bm{k} \down} -  V' \cos k_y (c_{\bm{k} \down}^{\dagger} f_{\bm{k} \down} + f_{\bm{k} \down}^{\dagger} c_{\bm{k} \down}) )$ with spin axis in the $x$-direction when an electric field is applied.
This quantity can be calculated by replacing the velocity operator $J_i$ in Eq.~(\ref{EQnonlinear}) with $J_{i \up (\down)}$ (but not changing $J_j$ and $J_k$ in Eq.~(\ref{EQnonlinear})). We call these spin-dependent conductivities $\sigma_{yyy \up}$ and $\sigma_{yyy \down}$. 
We see that the nonlinear conductivity for the $x$-up electrons is almost zero and the current is generated almost only by the the $x$-down electrons, as shown in Fig.~\ref{spinpolar}. Only at small interaction strengths, where the gap in the $x$-up DOS closes,  do $x$-up electrons contribute to the nonlinear conductivity. 
Thus, we find a nonlinear conductivity perpendicular to the magnetization of the system, which for strong interaction strengths is exclusively carried by one spin direction. As the spin-selective gap in the DOS is created by correlation effects for a wide range of parameters and does not need fine-tuning, we find a spin-selective nonreciprocal conductivity that is controlled by the direction of the magnetization of the material. 

\begin{figure}
\begin{tabular}{ll}
\includegraphics[width=4.5cm]{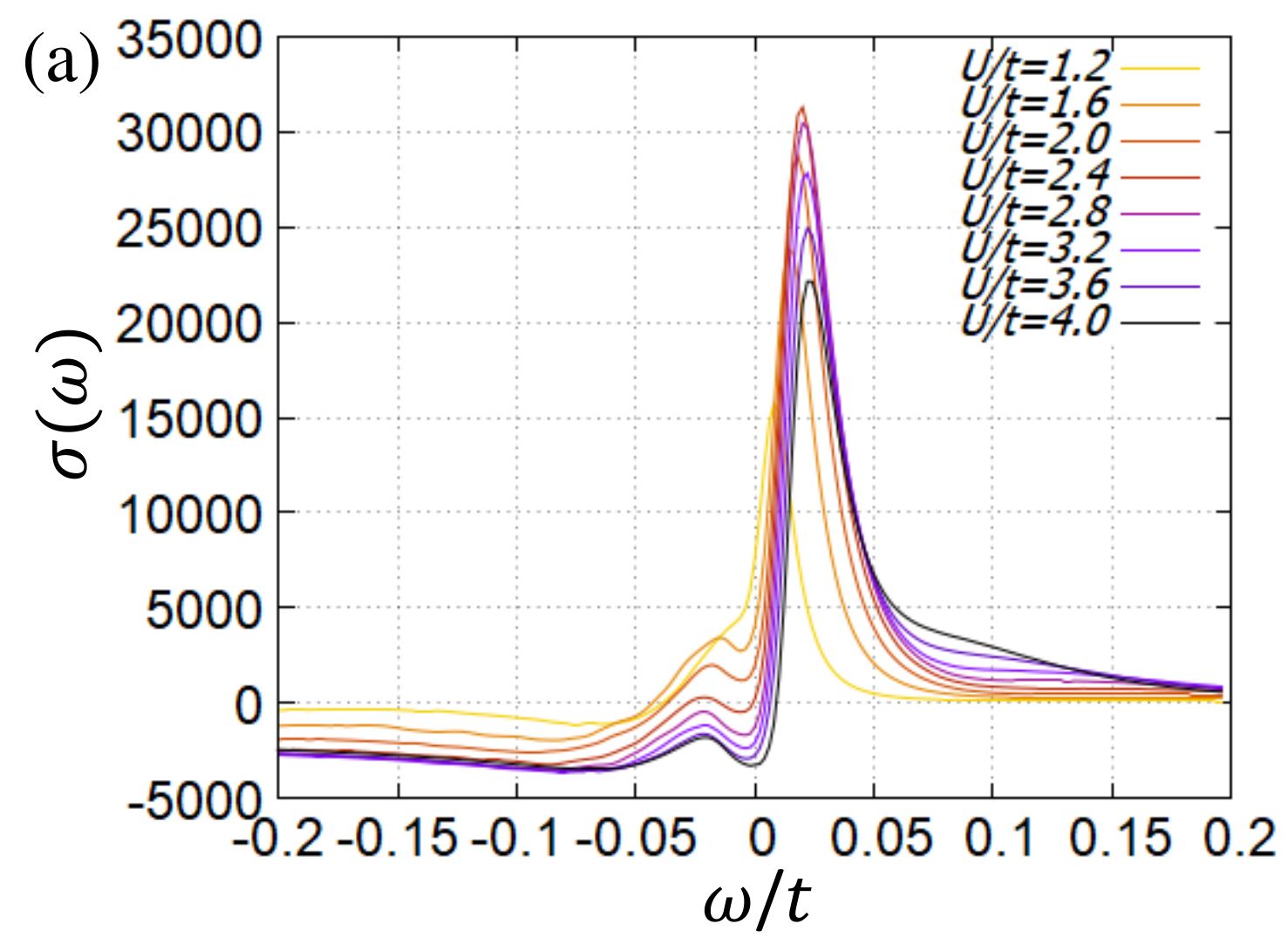}
&
\includegraphics[width=4.5cm]{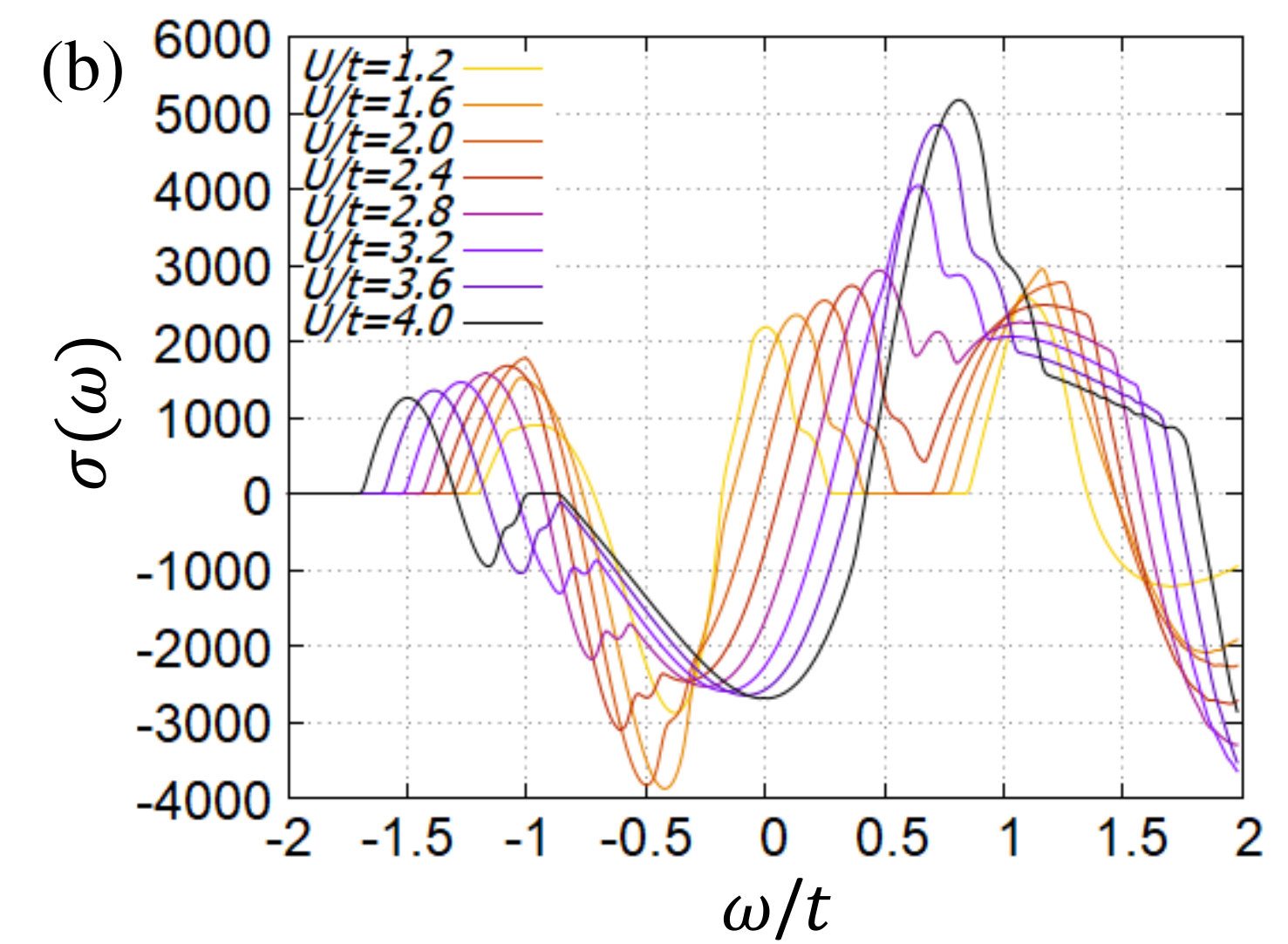}
\end{tabular}
\caption{\label{sign_U}(a) Frequency dependent contribution to the nonlinear conductivity at $T/t=0.2 \times 10^{-2}$. $\sigma (\omega)$ is defined via $\sigma_{yyy} = \int (\partial f(\omega) / \partial \omega) \sigma(\omega) d\omega$. The nonlinear conductivity in (a) is calculated using the full self-energy. Panel (b) shows the frequency dependent contribution to the nonlinear conductivity, $\sigma (\omega)$, calculated using only the magnetic shift, $\mathrm{Re} \Sigma (\omega=0)$. Parameters are the same as in Fig.~\ref{mag}.}
\end{figure}
\begin{figure*}[t]
\begin{tabular}{cccc}
\includegraphics[width=4.5cm]{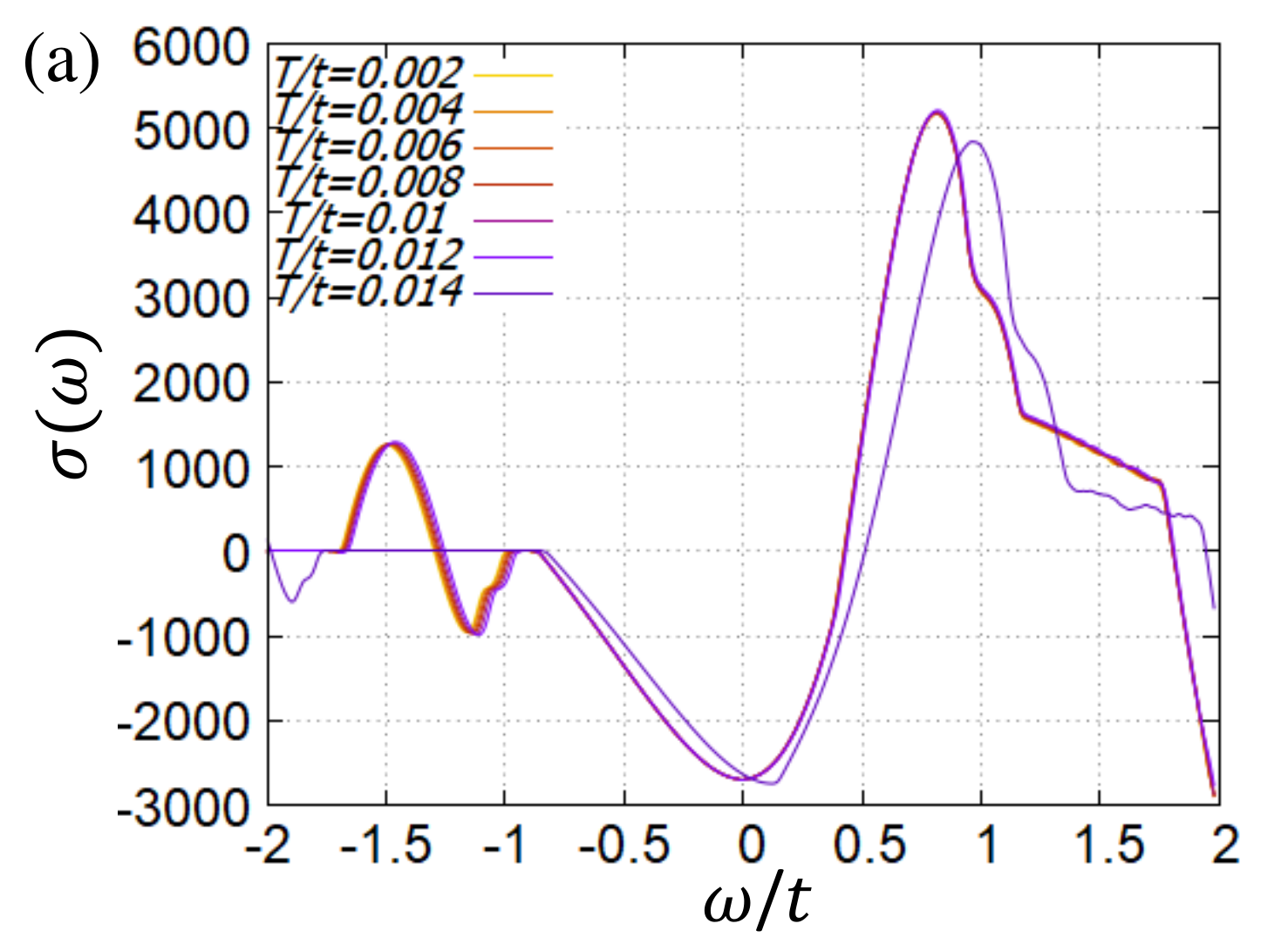}
&
\includegraphics[width=4.5cm]{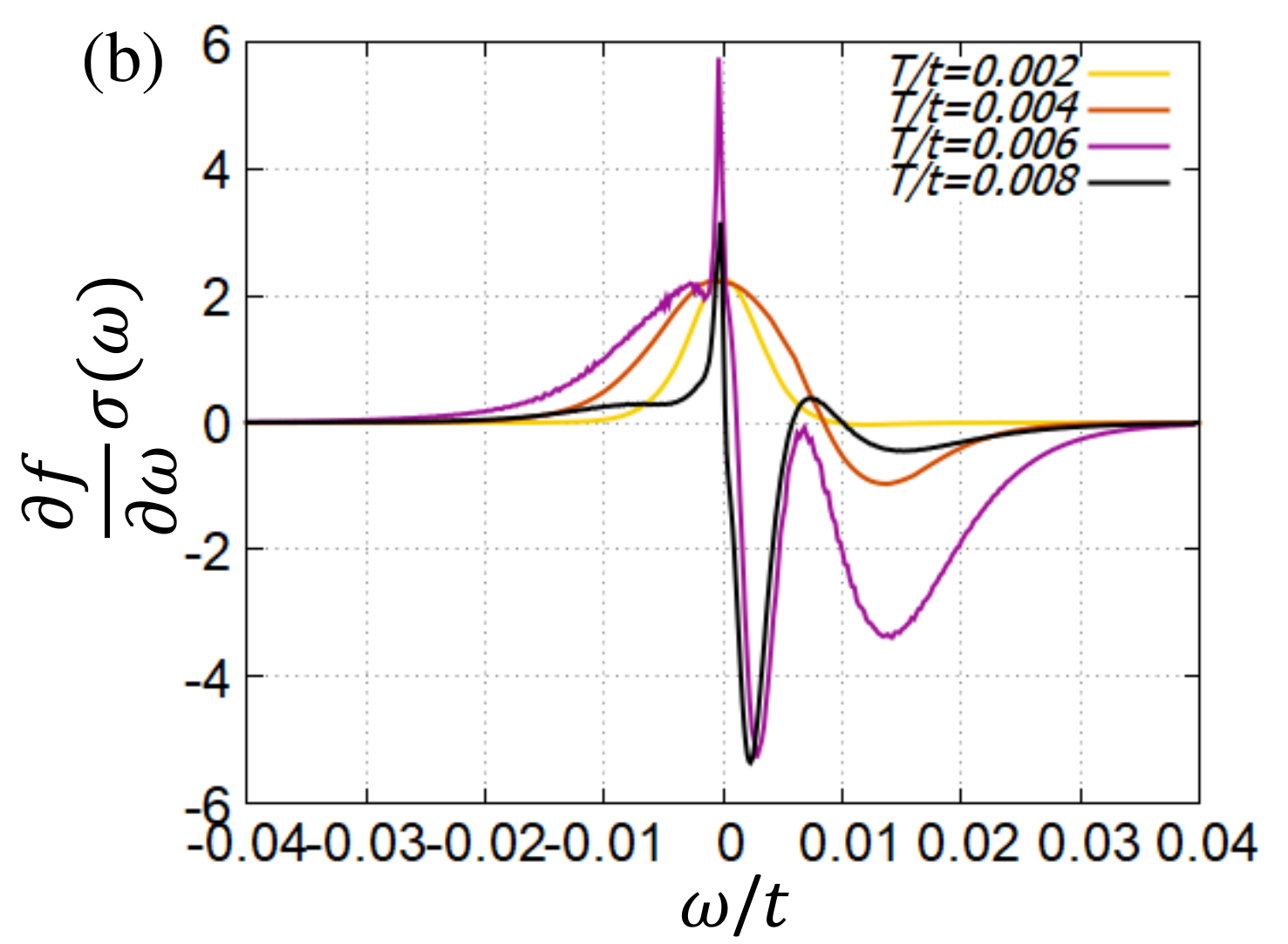}
&
\includegraphics[width=4.5cm]{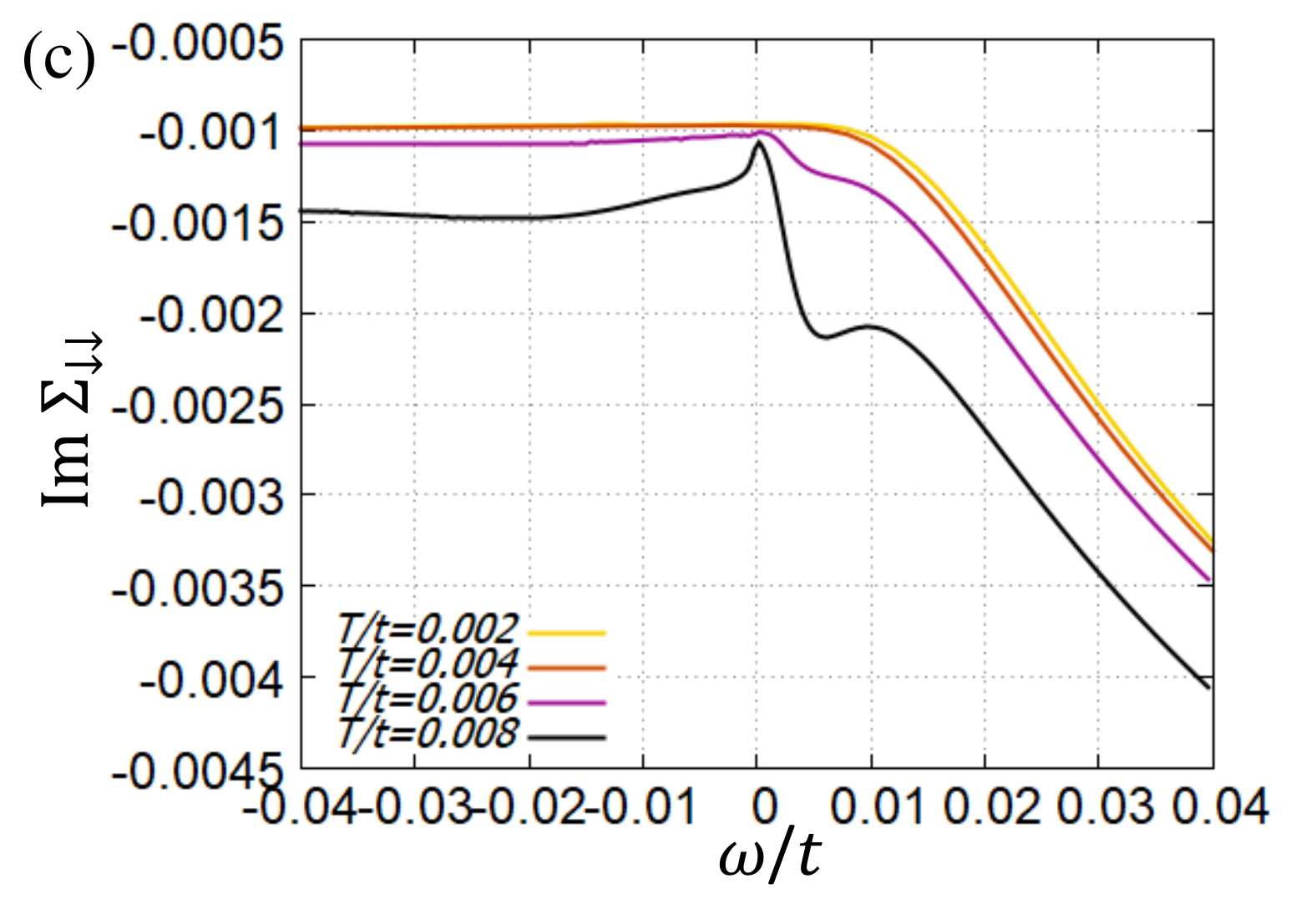}
&
\includegraphics[width=4.5cm]{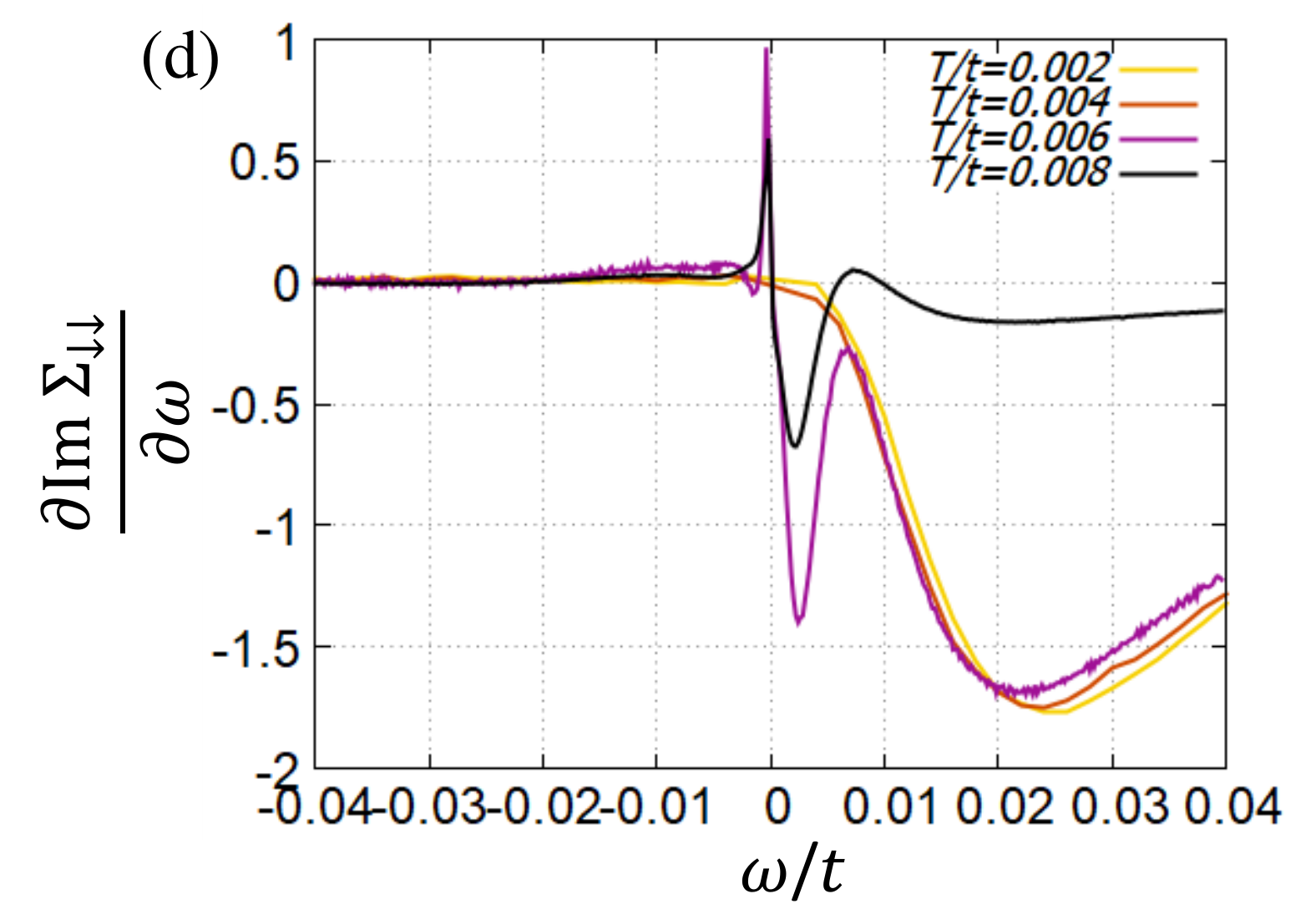}
\end{tabular}
\caption{\label{sign_U_2.0} (a) 
Frequency dependent contribution to the nonlinear conductivity for $U/t=4.0$ calculated using only the magnetic shift, $\mathrm{Re} \Sigma (\omega=0)$. $\sigma (\omega)$ is defined via $\sigma_{yyy} = \int (\partial f(\omega) / \partial \omega) \sigma(\omega) d\omega$.
Panel (b) shows $(\partial f(\omega) / \partial \omega) \sigma(\omega)$ calculated using the full self-energy for different temperatures. The data at $T/t=0.8 \times 10^{-2}$ is scaled by $1/10$. Panel (c) and (d) show the imaginary part of the self-energy for the down-spin (c), and the derivative of it (d). Parameters are the same as in Fig.~\ref{mag}. }
\end{figure*}
As seen in Fig.~\ref{nonlinear}, the sign of the nonlinear conductivity changes depending on the temperature and the interaction strength.
To understand the mechanism of this sign change, we analyze the contribution from each $\omega$ to the nonlinear conductivity in Fig.~\ref{sign_U} at $T/t= 0.2 \times 10^{-2}$. The nonlinear conductivity is then given as 
$\sigma_{yyy} = \int (\partial f(\omega) / \partial \omega) \sigma(\omega) d\omega$. 
Fig.~\ref{sign_U}(a) shows $\sigma (\omega)$ at $T/t=0.2 \times 10^{-2}$ calculated using the full self-energy. We clearly see the sign inversion at the Fermi energy ($\omega = 0$) and additional peaks near the Fermi energy. 
To understand the change of $\sigma (\omega)$ at the Fermi energy, we show in Fig.~\ref{sign_U} (b) results for the noninteracting model with constant magnetic fields. When calculating the nonlinear conductivity in Fig.~\ref{sign_U}(b), we approximate the self-energy as $\Sigma(\omega)=\text{Re}\Sigma(0)$ instead of using the full self-energy. This corresponds to a static magnetic field. 
Even in this case, the sign change occurs at the Fermi energy. In particular, we see that peaks with positive weight move to the Fermi energy when decreasing the interaction strength. With decreasing interaction strength, the Kondo screening becomes stronger and weakens the magnetization. As described above, spectral weight of the $f$ electrons shifts to the Fermi energy, contributing to the nonlinear conductivity.
Thus, we understand that the sign change in the nonlinear conductivity at low temperatures results from the magnetic shift and the Kondo screening changing the DOS at the Fermi energy.

On the other hand, the sign inversion with increasing temperature from $T/t=0.2 \times 10^{-2}$ at $U/t=4.0$ cannot be explained by the magnetic shift, as shown in Fig.~\ref{sign_U_2.0}(a), where we again approximate the self-energy as $\Sigma (\omega)=\mathrm{Re} \Sigma (\omega=0)$.  Including only the magnetic shift at different temperatures, we see that the sign of the nonlinear conductivity at the Fermi energy is always negative.
Thus, as large interation strengths, different correlation effects lead to a sign change of the nonlinear conductivity. 
In Fig.~\ref{sign_U_2.0}(b), we show again the frequency-dependent contribution to the nonlinear conduction, $(\partial f(\omega) / \partial \omega) \sigma(\omega)$ using the full self-energy.
At $T/t=0.2 \times 10^{-2}$ the shape resembles a Lorentz function taking only positive values.
However, when increasing the temperature, peaks with negative weight appear, which contribute to the sign-inversion. 
We found that these peaks originate from the self-energy (Fig.~\ref{sign_U_2.0}(c)), and in particular from the derivative of the imaginary part of the self-energy, shown in Fig.~\ref{sign_U_2.0}(d). Comparing Figs.~\ref{sign_U_2.0}(b) and (d), we see that positive and negative peaks in the nonlinear conductivity and the derivative of the self-energy correspond to each other.
Because the equation for the nonlinear conductivity includes the derivative of the Green's function, $\frac{\partial G^R}{\partial \omega}=-G^R (1-\frac{\partial \Sigma (\omega) }{ \partial \omega}) G^R$, the derivative of the self-energy contributes to the nonlinear conductivity. 
When increasing the temperature at large interaction strength, the shape of the self-energy strongly changes. These changes in the self-energy and its derivative result in a sign change of the nonlinear conductivity.
Thus, the nonlinear conductivity reveals more details about the single-particle spectrum and self-energy than the linear conductivity and can be a valuable tool for analyzing correlation effects.

\section{Conclusions} \label{conclusion}
In summary, we have analyzed the linear and nonlinear DC conductivity in the ferromagnetic state of the periodic Anderson model with Rashba-like SOC. 
We have confirmed the existence of the ferromagnetic phase for low-filled $c$ electrons in the presence of Rashba SOC. 
As observed in previous calculations, we have found a gap in the DOS at the Fermi energy for one spin-direction due to the interplay of the Kondo effect and ferromagnetism. Interestingly, our calculations have revealed that the linear conductivity is strongly enhanced in the ferromagnetic state, which can be explained by a change of the Fermi surface of the $c$ electrons. Analyzing the nonlinear DC conductivity using the Boltzmann equation and Green's function method, we have found that a nonlinear conductivity only exists in the ferromagnetic state perpendicular to the magnetization of the system. Furthermore, due to the gap in the DOS of one spin direction, the linear and nonlinear conductivities are strongly spin-dependent. Thus, our calculations have revealed a  spin-dependent nonreciprocal conductivity depending on the direction of the magnetization of the system. We believe that such spin-dependent nonreciprocal conductivity might be useful for spintronic applications.
Furthermore, we have observed that the sign of the nonlinear conductivity changes depending on the temperature and the interaction strength. We have identified two different mechanisms responsible for this sign change. 
At low temperature, the magnetic energy shift and the change of the Fermi surface due to the Kondo effect result in a sign change when increasing the interaction strength. On the other hand, at strong interaction strengths, the changes of the self-energy when increasing the temperature result in a sign change of the nonlinear conductivity.  This result demonstrates that the nonlinear conductivity reveals much more details of the excitation spectrum of the material than the linear conductivity.

\section*{acknowledge}
K.S. is full of gratitude to Yoshihiro Michishita for his valuable advice and discussions. K.S. also thanks Kazuhiro Kimura, Akira Kofuji, Hikaru Watanabe, and Michiya Chazono for fruitful discussions. R.P. is supported by JSPS, KAKENHI Grant No. JP18K03511. The computer calculations in this work have been done using the facilities of the Supercomputer Center, the Institute for Solid
State Physics, the University of Tokyo.

\bibliography{reference.bib}

\begin{thebibliography}{62}%
\makeatletter
\providecommand \@ifxundefined [1]{%
 \@ifx{#1\undefined}
}%
\providecommand \@ifnum [1]{%
 \ifnum #1\expandafter \@firstoftwo
 \else \expandafter \@secondoftwo
 \fi
}%
\providecommand \@ifx [1]{%
 \ifx #1\expandafter \@firstoftwo
 \else \expandafter \@secondoftwo
 \fi
}%
\providecommand \natexlab [1]{#1}%
\providecommand \enquote  [1]{``#1''}%
\providecommand \bibnamefont  [1]{#1}%
\providecommand \bibfnamefont [1]{#1}%
\providecommand \citenamefont [1]{#1}%
\providecommand \href@noop [0]{\@secondoftwo}%
\providecommand \href [0]{\begingroup \@sanitize@url \@href}%
\providecommand \@href[1]{\@@startlink{#1}\@@href}%
\providecommand \@@href[1]{\endgroup#1\@@endlink}%
\providecommand \@sanitize@url [0]{\catcode `\\12\catcode `\$12\catcode
  `\&12\catcode `\#12\catcode `\^12\catcode `\_12\catcode `\%12\relax}%
\providecommand \@@startlink[1]{}%
\providecommand \@@endlink[0]{}%
\providecommand \url  [0]{\begingroup\@sanitize@url \@url }%
\providecommand \@url [1]{\endgroup\@href {#1}{\urlprefix }}%
\providecommand \urlprefix  [0]{URL }%
\providecommand \Eprint [0]{\href }%
\providecommand \doibase [0]{https://doi.org/}%
\providecommand \selectlanguage [0]{\@gobble}%
\providecommand \bibinfo  [0]{\@secondoftwo}%
\providecommand \bibfield  [0]{\@secondoftwo}%
\providecommand \translation [1]{[#1]}%
\providecommand \BibitemOpen [0]{}%
\providecommand \bibitemStop [0]{}%
\providecommand \bibitemNoStop [0]{.\EOS\space}%
\providecommand \EOS [0]{\spacefactor3000\relax}%
\providecommand \BibitemShut  [1]{\csname bibitem#1\endcsname}%
\let\auto@bib@innerbib\@empty
\bibitem [{\citenamefont {Witczak-Krempa}\ \emph {et~al.}(2014)\citenamefont
  {Witczak-Krempa}, \citenamefont {Chen}, \citenamefont {Kim},\ and\
  \citenamefont {Balents}}]{annurev-conmatphys-020911-125138}%
  \BibitemOpen
  \bibfield  {author} {\bibinfo {author} {\bibfnamefont {W.}~\bibnamefont
  {Witczak-Krempa}}, \bibinfo {author} {\bibfnamefont {G.}~\bibnamefont
  {Chen}}, \bibinfo {author} {\bibfnamefont {Y.~B.}\ \bibnamefont {Kim}},\ and\
  \bibinfo {author} {\bibfnamefont {L.}~\bibnamefont {Balents}},\ }\bibfield
  {title} {\bibinfo {title} {Correlated quantum phenomena in the strong
  spin-orbit regime},\ }\href
  {https://doi.org/10.1146/annurev-conmatphys-020911-125138} {\bibfield
  {journal} {\bibinfo  {journal} {Annual Review of Condensed Matter Physics}\
  }\textbf {\bibinfo {volume} {5}},\ \bibinfo {pages} {57} (\bibinfo {year}
  {2014})},\ \Eprint
  {https://arxiv.org/abs/https://doi.org/10.1146/annurev-conmatphys-020911-125138}
  {https://doi.org/10.1146/annurev-conmatphys-020911-125138} \BibitemShut
  {NoStop}%
\bibitem [{\citenamefont {Manchon}\ \emph {et~al.}(2015)\citenamefont
  {Manchon}, \citenamefont {Koo}, \citenamefont {Nitta}, \citenamefont
  {Frolov},\ and\ \citenamefont {Duine}}]{Manchon2015}%
  \BibitemOpen
  \bibfield  {author} {\bibinfo {author} {\bibfnamefont {A.}~\bibnamefont
  {Manchon}}, \bibinfo {author} {\bibfnamefont {H.~C.}\ \bibnamefont {Koo}},
  \bibinfo {author} {\bibfnamefont {J.}~\bibnamefont {Nitta}}, \bibinfo
  {author} {\bibfnamefont {S.~M.}\ \bibnamefont {Frolov}},\ and\ \bibinfo
  {author} {\bibfnamefont {R.~A.}\ \bibnamefont {Duine}},\ }\bibfield  {title}
  {\bibinfo {title} {New perspectives for rashba spin--orbit coupling},\ }\href
  {https://doi.org/10.1038/nmat4360} {\bibfield  {journal} {\bibinfo  {journal}
  {Nature Materials}\ }\textbf {\bibinfo {volume} {14}},\ \bibinfo {pages}
  {871} (\bibinfo {year} {2015})}\BibitemShut {NoStop}%
\bibitem [{\citenamefont {Smidman}\ \emph {et~al.}(2017)\citenamefont
  {Smidman}, \citenamefont {Salamon}, \citenamefont {Yuan},\ and\ \citenamefont
  {Agterberg}}]{Smidman_2017}%
  \BibitemOpen
  \bibfield  {author} {\bibinfo {author} {\bibfnamefont {M.}~\bibnamefont
  {Smidman}}, \bibinfo {author} {\bibfnamefont {M.~B.}\ \bibnamefont
  {Salamon}}, \bibinfo {author} {\bibfnamefont {H.~Q.}\ \bibnamefont {Yuan}},\
  and\ \bibinfo {author} {\bibfnamefont {D.~F.}\ \bibnamefont {Agterberg}},\
  }\bibfield  {title} {\bibinfo {title} {Superconductivity and
  spin{\textendash}orbit coupling in non-centrosymmetric materials: a review},\
  }\href {https://doi.org/10.1088/1361-6633/80/3/036501} {\bibfield  {journal}
  {\bibinfo  {journal} {Reports on Progress in Physics}\ }\textbf {\bibinfo
  {volume} {80}},\ \bibinfo {pages} {036501} (\bibinfo {year}
  {2017})}\BibitemShut {NoStop}%
\bibitem [{\citenamefont {Goh}\ \emph {et~al.}(2012)\citenamefont {Goh},
  \citenamefont {Mizukami}, \citenamefont {Shishido}, \citenamefont {Watanabe},
  \citenamefont {Yasumoto}, \citenamefont {Shimozawa}, \citenamefont
  {Yamashita}, \citenamefont {Terashima}, \citenamefont {Yanase}, \citenamefont
  {Shibauchi}, \citenamefont {Buzdin},\ and\ \citenamefont
  {Matsuda}}]{PhysRevLett.109.157006}%
  \BibitemOpen
  \bibfield  {author} {\bibinfo {author} {\bibfnamefont {S.~K.}\ \bibnamefont
  {Goh}}, \bibinfo {author} {\bibfnamefont {Y.}~\bibnamefont {Mizukami}},
  \bibinfo {author} {\bibfnamefont {H.}~\bibnamefont {Shishido}}, \bibinfo
  {author} {\bibfnamefont {D.}~\bibnamefont {Watanabe}}, \bibinfo {author}
  {\bibfnamefont {S.}~\bibnamefont {Yasumoto}}, \bibinfo {author}
  {\bibfnamefont {M.}~\bibnamefont {Shimozawa}}, \bibinfo {author}
  {\bibfnamefont {M.}~\bibnamefont {Yamashita}}, \bibinfo {author}
  {\bibfnamefont {T.}~\bibnamefont {Terashima}}, \bibinfo {author}
  {\bibfnamefont {Y.}~\bibnamefont {Yanase}}, \bibinfo {author} {\bibfnamefont
  {T.}~\bibnamefont {Shibauchi}}, \bibinfo {author} {\bibfnamefont {A.~I.}\
  \bibnamefont {Buzdin}},\ and\ \bibinfo {author} {\bibfnamefont
  {Y.}~\bibnamefont {Matsuda}},\ }\bibfield  {title} {\bibinfo {title}
  {Anomalous upper critical field in
  ${\mathrm{cecoin}}_{5}/{\mathrm{ybcoin}}_{5}$ superlattices with a
  rashba-type heavy fermion interface},\ }\href
  {https://doi.org/10.1103/PhysRevLett.109.157006} {\bibfield  {journal}
  {\bibinfo  {journal} {Phys. Rev. Lett.}\ }\textbf {\bibinfo {volume} {109}},\
  \bibinfo {pages} {157006} (\bibinfo {year} {2012})}\BibitemShut {NoStop}%
\bibitem [{\citenamefont {Shimozawa}\ \emph {et~al.}(2014)\citenamefont
  {Shimozawa}, \citenamefont {Goh}, \citenamefont {Endo}, \citenamefont
  {Kobayashi}, \citenamefont {Watashige}, \citenamefont {Mizukami},
  \citenamefont {Ikeda}, \citenamefont {Shishido}, \citenamefont {Yanase},
  \citenamefont {Terashima}, \citenamefont {Shibauchi},\ and\ \citenamefont
  {Matsuda}}]{PhysRevLett.112.156404}%
  \BibitemOpen
  \bibfield  {author} {\bibinfo {author} {\bibfnamefont {M.}~\bibnamefont
  {Shimozawa}}, \bibinfo {author} {\bibfnamefont {S.~K.}\ \bibnamefont {Goh}},
  \bibinfo {author} {\bibfnamefont {R.}~\bibnamefont {Endo}}, \bibinfo {author}
  {\bibfnamefont {R.}~\bibnamefont {Kobayashi}}, \bibinfo {author}
  {\bibfnamefont {T.}~\bibnamefont {Watashige}}, \bibinfo {author}
  {\bibfnamefont {Y.}~\bibnamefont {Mizukami}}, \bibinfo {author}
  {\bibfnamefont {H.}~\bibnamefont {Ikeda}}, \bibinfo {author} {\bibfnamefont
  {H.}~\bibnamefont {Shishido}}, \bibinfo {author} {\bibfnamefont
  {Y.}~\bibnamefont {Yanase}}, \bibinfo {author} {\bibfnamefont
  {T.}~\bibnamefont {Terashima}}, \bibinfo {author} {\bibfnamefont
  {T.}~\bibnamefont {Shibauchi}},\ and\ \bibinfo {author} {\bibfnamefont
  {Y.}~\bibnamefont {Matsuda}},\ }\bibfield  {title} {\bibinfo {title}
  {Controllable rashba spin-orbit interaction in artificially engineered
  superlattices involving the heavy-fermion superconductor
  ${\mathrm{cecoin}}_{5}$},\ }\href
  {https://doi.org/10.1103/PhysRevLett.112.156404} {\bibfield  {journal}
  {\bibinfo  {journal} {Phys. Rev. Lett.}\ }\textbf {\bibinfo {volume} {112}},\
  \bibinfo {pages} {156404} (\bibinfo {year} {2014})}\BibitemShut {NoStop}%
\bibitem [{\citenamefont {Lin}\ \emph {et~al.}(2019)\citenamefont {Lin},
  \citenamefont {Li}, \citenamefont {Do{\u{g}}an}, \citenamefont {Li},
  \citenamefont {Rotella}, \citenamefont {Yu}, \citenamefont {Zhang},
  \citenamefont {Li}, \citenamefont {Lew}, \citenamefont {Wang}, \citenamefont
  {Prellier}, \citenamefont {Pennycook}, \citenamefont {Chen}, \citenamefont
  {Zhong}, \citenamefont {Manchon},\ and\ \citenamefont {Wu}}]{Lin2019}%
  \BibitemOpen
  \bibfield  {author} {\bibinfo {author} {\bibfnamefont {W.}~\bibnamefont
  {Lin}}, \bibinfo {author} {\bibfnamefont {L.}~\bibnamefont {Li}}, \bibinfo
  {author} {\bibfnamefont {F.}~\bibnamefont {Do{\u{g}}an}}, \bibinfo {author}
  {\bibfnamefont {C.}~\bibnamefont {Li}}, \bibinfo {author} {\bibfnamefont
  {H.}~\bibnamefont {Rotella}}, \bibinfo {author} {\bibfnamefont
  {X.}~\bibnamefont {Yu}}, \bibinfo {author} {\bibfnamefont {B.}~\bibnamefont
  {Zhang}}, \bibinfo {author} {\bibfnamefont {Y.}~\bibnamefont {Li}}, \bibinfo
  {author} {\bibfnamefont {W.~S.}\ \bibnamefont {Lew}}, \bibinfo {author}
  {\bibfnamefont {S.}~\bibnamefont {Wang}}, \bibinfo {author} {\bibfnamefont
  {W.}~\bibnamefont {Prellier}}, \bibinfo {author} {\bibfnamefont {S.~J.}\
  \bibnamefont {Pennycook}}, \bibinfo {author} {\bibfnamefont {J.}~\bibnamefont
  {Chen}}, \bibinfo {author} {\bibfnamefont {Z.}~\bibnamefont {Zhong}},
  \bibinfo {author} {\bibfnamefont {A.}~\bibnamefont {Manchon}},\ and\ \bibinfo
  {author} {\bibfnamefont {T.}~\bibnamefont {Wu}},\ }\bibfield  {title}
  {\bibinfo {title} {Interface-based tuning of rashba spin-orbit interaction in
  asymmetric oxide heterostructures with 3d electrons},\ }\href
  {https://doi.org/10.1038/s41467-019-10961-z} {\bibfield  {journal} {\bibinfo
  {journal} {Nature Communications}\ }\textbf {\bibinfo {volume} {10}},\
  \bibinfo {pages} {3052} (\bibinfo {year} {2019})}\BibitemShut {NoStop}%
\bibitem [{\citenamefont {Peters}\ and\ \citenamefont
  {Yanase}(2018)}]{PhysRevB.97.115128}%
  \BibitemOpen
  \bibfield  {author} {\bibinfo {author} {\bibfnamefont {R.}~\bibnamefont
  {Peters}}\ and\ \bibinfo {author} {\bibfnamefont {Y.}~\bibnamefont
  {Yanase}},\ }\bibfield  {title} {\bibinfo {title} {Strong enhancement of the
  edelstein effect in $f$-electron systems},\ }\href
  {https://doi.org/10.1103/PhysRevB.97.115128} {\bibfield  {journal} {\bibinfo
  {journal} {Phys. Rev. B}\ }\textbf {\bibinfo {volume} {97}},\ \bibinfo
  {pages} {115128} (\bibinfo {year} {2018})}\BibitemShut {NoStop}%
\bibitem [{\citenamefont {Chernyshov}\ \emph {et~al.}(2009)\citenamefont
  {Chernyshov}, \citenamefont {Overby}, \citenamefont {Liu}, \citenamefont
  {Furdyna}, \citenamefont {Lyanda-Geller},\ and\ \citenamefont
  {Rokhinson}}]{Chernyshov2009}%
  \BibitemOpen
  \bibfield  {author} {\bibinfo {author} {\bibfnamefont {A.}~\bibnamefont
  {Chernyshov}}, \bibinfo {author} {\bibfnamefont {M.}~\bibnamefont {Overby}},
  \bibinfo {author} {\bibfnamefont {X.}~\bibnamefont {Liu}}, \bibinfo {author}
  {\bibfnamefont {J.~K.}\ \bibnamefont {Furdyna}}, \bibinfo {author}
  {\bibfnamefont {Y.}~\bibnamefont {Lyanda-Geller}},\ and\ \bibinfo {author}
  {\bibfnamefont {L.~P.}\ \bibnamefont {Rokhinson}},\ }\bibfield  {title}
  {\bibinfo {title} {Evidence for reversible control of magnetization in a
  ferromagnetic material by means of spin--orbit magnetic field},\ }\href
  {https://doi.org/10.1038/nphys1362} {\bibfield  {journal} {\bibinfo
  {journal} {Nature Physics}\ }\textbf {\bibinfo {volume} {5}},\ \bibinfo
  {pages} {656} (\bibinfo {year} {2009})}\BibitemShut {NoStop}%
\bibitem [{\citenamefont {Edelstein}(1990)}]{EDELSTEIN1990233}%
  \BibitemOpen
  \bibfield  {author} {\bibinfo {author} {\bibfnamefont {V.}~\bibnamefont
  {Edelstein}},\ }\bibfield  {title} {\bibinfo {title} {Spin polarization of
  conduction electrons induced by electric current in two-dimensional
  asymmetric electron systems},\ }\href
  {https://doi.org/https://doi.org/10.1016/0038-1098(90)90963-C} {\bibfield
  {journal} {\bibinfo  {journal} {Solid State Communications}\ }\textbf
  {\bibinfo {volume} {73}},\ \bibinfo {pages} {233} (\bibinfo {year}
  {1990})}\BibitemShut {NoStop}%
\bibitem [{\citenamefont {Hasan}\ and\ \citenamefont {Kane}(2010)}]{Hasan2010}%
  \BibitemOpen
  \bibfield  {author} {\bibinfo {author} {\bibfnamefont {M.~Z.}\ \bibnamefont
  {Hasan}}\ and\ \bibinfo {author} {\bibfnamefont {C.~L.}\ \bibnamefont
  {Kane}},\ }\bibfield  {title} {\bibinfo {title} {Colloquium: Topological
  insulators},\ }\href {https://doi.org/10.1103/RevModPhys.82.3045} {\bibfield
  {journal} {\bibinfo  {journal} {Rev. Mod. Phys.}\ }\textbf {\bibinfo {volume}
  {82}},\ \bibinfo {pages} {3045} (\bibinfo {year} {2010})}\BibitemShut
  {NoStop}%
\bibitem [{\citenamefont {Rikken}\ \emph {et~al.}(2001)\citenamefont {Rikken},
  \citenamefont {F\"olling},\ and\ \citenamefont {Wyder}}]{Rikken2001}%
  \BibitemOpen
  \bibfield  {author} {\bibinfo {author} {\bibfnamefont {G.~L. J.~A.}\
  \bibnamefont {Rikken}}, \bibinfo {author} {\bibfnamefont {J.}~\bibnamefont
  {F\"olling}},\ and\ \bibinfo {author} {\bibfnamefont {P.}~\bibnamefont
  {Wyder}},\ }\bibfield  {title} {\bibinfo {title} {Electrical magnetochiral
  anisotropy},\ }\href {https://doi.org/10.1103/PhysRevLett.87.236602}
  {\bibfield  {journal} {\bibinfo  {journal} {Phys. Rev. Lett.}\ }\textbf
  {\bibinfo {volume} {87}},\ \bibinfo {pages} {236602} (\bibinfo {year}
  {2001})}\BibitemShut {NoStop}%
\bibitem [{\citenamefont {Ishizaka}\ \emph {et~al.}(2011)\citenamefont
  {Ishizaka}, \citenamefont {Bahramy}, \citenamefont {Murakawa}, \citenamefont
  {Sakano}, \citenamefont {Shimojima}, \citenamefont {Sonobe}, \citenamefont
  {Koizumi}, \citenamefont {Shin}, \citenamefont {Miyahara}, \citenamefont
  {Kimura}, \citenamefont {Miyamoto}, \citenamefont {Okuda}, \citenamefont
  {Namatame}, \citenamefont {Taniguchi}, \citenamefont {Arita}, \citenamefont
  {Nagaosa}, \citenamefont {Kobayashi}, \citenamefont {Murakami}, \citenamefont
  {Kumai}, \citenamefont {Kaneko}, \citenamefont {Onose},\ and\ \citenamefont
  {Tokura}}]{Ishizaka2011}%
  \BibitemOpen
  \bibfield  {author} {\bibinfo {author} {\bibfnamefont {K.}~\bibnamefont
  {Ishizaka}}, \bibinfo {author} {\bibfnamefont {M.~S.}\ \bibnamefont
  {Bahramy}}, \bibinfo {author} {\bibfnamefont {H.}~\bibnamefont {Murakawa}},
  \bibinfo {author} {\bibfnamefont {M.}~\bibnamefont {Sakano}}, \bibinfo
  {author} {\bibfnamefont {T.}~\bibnamefont {Shimojima}}, \bibinfo {author}
  {\bibfnamefont {T.}~\bibnamefont {Sonobe}}, \bibinfo {author} {\bibfnamefont
  {K.}~\bibnamefont {Koizumi}}, \bibinfo {author} {\bibfnamefont
  {S.}~\bibnamefont {Shin}}, \bibinfo {author} {\bibfnamefont {H.}~\bibnamefont
  {Miyahara}}, \bibinfo {author} {\bibfnamefont {A.}~\bibnamefont {Kimura}},
  \bibinfo {author} {\bibfnamefont {K.}~\bibnamefont {Miyamoto}}, \bibinfo
  {author} {\bibfnamefont {T.}~\bibnamefont {Okuda}}, \bibinfo {author}
  {\bibfnamefont {H.}~\bibnamefont {Namatame}}, \bibinfo {author}
  {\bibfnamefont {M.}~\bibnamefont {Taniguchi}}, \bibinfo {author}
  {\bibfnamefont {R.}~\bibnamefont {Arita}}, \bibinfo {author} {\bibfnamefont
  {N.}~\bibnamefont {Nagaosa}}, \bibinfo {author} {\bibfnamefont
  {K.}~\bibnamefont {Kobayashi}}, \bibinfo {author} {\bibfnamefont
  {Y.}~\bibnamefont {Murakami}}, \bibinfo {author} {\bibfnamefont
  {R.}~\bibnamefont {Kumai}}, \bibinfo {author} {\bibfnamefont
  {Y.}~\bibnamefont {Kaneko}}, \bibinfo {author} {\bibfnamefont
  {Y.}~\bibnamefont {Onose}},\ and\ \bibinfo {author} {\bibfnamefont
  {Y.}~\bibnamefont {Tokura}},\ }\bibfield  {title} {\bibinfo {title} {Giant
  rashba-type spin splitting in bulk bitei},\ }\href
  {https://doi.org/10.1038/nmat3051} {\bibfield  {journal} {\bibinfo  {journal}
  {Nature Materials}\ }\textbf {\bibinfo {volume} {10}},\ \bibinfo {pages}
  {521} (\bibinfo {year} {2011})}\BibitemShut {NoStop}%
\bibitem [{\citenamefont {Rikken}\ and\ \citenamefont
  {Wyder}(2005)}]{Rikken2015}%
  \BibitemOpen
  \bibfield  {author} {\bibinfo {author} {\bibfnamefont {G.~L. J.~A.}\
  \bibnamefont {Rikken}}\ and\ \bibinfo {author} {\bibfnamefont
  {P.}~\bibnamefont {Wyder}},\ }\bibfield  {title} {\bibinfo {title}
  {Magnetoelectric anisotropy in diffusive transport},\ }\href
  {https://doi.org/10.1103/PhysRevLett.94.016601} {\bibfield  {journal}
  {\bibinfo  {journal} {Phys. Rev. Lett.}\ }\textbf {\bibinfo {volume} {94}},\
  \bibinfo {pages} {016601} (\bibinfo {year} {2005})}\BibitemShut {NoStop}%
\bibitem [{\citenamefont {Avci}\ \emph
  {et~al.}(2015{\natexlab{a}})\citenamefont {Avci}, \citenamefont {Garello},
  \citenamefont {Ghosh}, \citenamefont {Gabureac}, \citenamefont {Alvarado},\
  and\ \citenamefont {Gambardella}}]{Avci2015}%
  \BibitemOpen
  \bibfield  {author} {\bibinfo {author} {\bibfnamefont {C.~O.}\ \bibnamefont
  {Avci}}, \bibinfo {author} {\bibfnamefont {K.}~\bibnamefont {Garello}},
  \bibinfo {author} {\bibfnamefont {A.}~\bibnamefont {Ghosh}}, \bibinfo
  {author} {\bibfnamefont {M.}~\bibnamefont {Gabureac}}, \bibinfo {author}
  {\bibfnamefont {S.~F.}\ \bibnamefont {Alvarado}},\ and\ \bibinfo {author}
  {\bibfnamefont {P.}~\bibnamefont {Gambardella}},\ }\bibfield  {title}
  {\bibinfo {title} {Unidirectional spin hall magnetoresistance in
  ferromagnet/normal metal bilayers},\ }\href
  {https://doi.org/10.1038/nphys3356} {\bibfield  {journal} {\bibinfo
  {journal} {Nature Physics}\ }\textbf {\bibinfo {volume} {11}},\ \bibinfo
  {pages} {570} (\bibinfo {year} {2015}{\natexlab{a}})}\BibitemShut {NoStop}%
\bibitem [{\citenamefont {Avci}\ \emph
  {et~al.}(2015{\natexlab{b}})\citenamefont {Avci}, \citenamefont {Garello},
  \citenamefont {Mendil}, \citenamefont {Ghosh}, \citenamefont {Blasakis},
  \citenamefont {Gabureac}, \citenamefont {Trassin}, \citenamefont {Fiebig},\
  and\ \citenamefont {Gambardella}}]{Avci20152}%
  \BibitemOpen
  \bibfield  {author} {\bibinfo {author} {\bibfnamefont {C.~O.}\ \bibnamefont
  {Avci}}, \bibinfo {author} {\bibfnamefont {K.}~\bibnamefont {Garello}},
  \bibinfo {author} {\bibfnamefont {J.}~\bibnamefont {Mendil}}, \bibinfo
  {author} {\bibfnamefont {A.}~\bibnamefont {Ghosh}}, \bibinfo {author}
  {\bibfnamefont {N.}~\bibnamefont {Blasakis}}, \bibinfo {author}
  {\bibfnamefont {M.}~\bibnamefont {Gabureac}}, \bibinfo {author}
  {\bibfnamefont {M.}~\bibnamefont {Trassin}}, \bibinfo {author} {\bibfnamefont
  {M.}~\bibnamefont {Fiebig}},\ and\ \bibinfo {author} {\bibfnamefont
  {P.}~\bibnamefont {Gambardella}},\ }\bibfield  {title} {\bibinfo {title}
  {Magnetoresistance of heavy and light metal/ferromagnet bilayers},\ }\href
  {https://doi.org/10.1063/1.4935497} {\bibfield  {journal} {\bibinfo
  {journal} {Applied Physics Letters}\ }\textbf {\bibinfo {volume} {107}},\
  \bibinfo {pages} {192405} (\bibinfo {year} {2015}{\natexlab{b}})},\ \Eprint
  {https://arxiv.org/abs/https://doi.org/10.1063/1.4935497}
  {https://doi.org/10.1063/1.4935497} \BibitemShut {NoStop}%
\bibitem [{\citenamefont {Olejn\'{\i}k}\ \emph {et~al.}(2015)\citenamefont
  {Olejn\'{\i}k}, \citenamefont {Nov\'ak}, \citenamefont {Wunderlich},\ and\
  \citenamefont {Jungwirth}}]{Olejn2015}%
  \BibitemOpen
  \bibfield  {author} {\bibinfo {author} {\bibfnamefont {K.}~\bibnamefont
  {Olejn\'{\i}k}}, \bibinfo {author} {\bibfnamefont {V.}~\bibnamefont
  {Nov\'ak}}, \bibinfo {author} {\bibfnamefont {J.}~\bibnamefont
  {Wunderlich}},\ and\ \bibinfo {author} {\bibfnamefont {T.}~\bibnamefont
  {Jungwirth}},\ }\bibfield  {title} {\bibinfo {title} {Electrical detection of
  magnetization reversal without auxiliary magnets},\ }\href
  {https://doi.org/10.1103/PhysRevB.91.180402} {\bibfield  {journal} {\bibinfo
  {journal} {Phys. Rev. B}\ }\textbf {\bibinfo {volume} {91}},\ \bibinfo
  {pages} {180402} (\bibinfo {year} {2015})}\BibitemShut {NoStop}%
\bibitem [{\citenamefont {Ideue}\ \emph {et~al.}(2017)\citenamefont {Ideue},
  \citenamefont {Hamamoto}, \citenamefont {Koshikawa}, \citenamefont {Ezawa},
  \citenamefont {Shimizu}, \citenamefont {Kaneko}, \citenamefont {Tokura},
  \citenamefont {Nagaosa},\ and\ \citenamefont {Iwasa}}]{Ideue2017}%
  \BibitemOpen
  \bibfield  {author} {\bibinfo {author} {\bibfnamefont {T.}~\bibnamefont
  {Ideue}}, \bibinfo {author} {\bibfnamefont {K.}~\bibnamefont {Hamamoto}},
  \bibinfo {author} {\bibfnamefont {S.}~\bibnamefont {Koshikawa}}, \bibinfo
  {author} {\bibfnamefont {M.}~\bibnamefont {Ezawa}}, \bibinfo {author}
  {\bibfnamefont {S.}~\bibnamefont {Shimizu}}, \bibinfo {author} {\bibfnamefont
  {Y.}~\bibnamefont {Kaneko}}, \bibinfo {author} {\bibfnamefont
  {Y.}~\bibnamefont {Tokura}}, \bibinfo {author} {\bibfnamefont
  {N.}~\bibnamefont {Nagaosa}},\ and\ \bibinfo {author} {\bibfnamefont
  {Y.}~\bibnamefont {Iwasa}},\ }\bibfield  {title} {\bibinfo {title} {Bulk
  rectification effect in a polar semiconductor},\ }\href
  {https://doi.org/10.1038/nphys4056} {\bibfield  {journal} {\bibinfo
  {journal} {Nature Physics}\ }\textbf {\bibinfo {volume} {13}},\ \bibinfo
  {pages} {578} (\bibinfo {year} {2017})}\BibitemShut {NoStop}%
\bibitem [{\citenamefont {Yasuda}\ \emph {et~al.}(2016)\citenamefont {Yasuda},
  \citenamefont {Tsukazaki}, \citenamefont {Yoshimi}, \citenamefont
  {Takahashi}, \citenamefont {Kawasaki},\ and\ \citenamefont
  {Tokura}}]{Yasuda2016}%
  \BibitemOpen
  \bibfield  {author} {\bibinfo {author} {\bibfnamefont {K.}~\bibnamefont
  {Yasuda}}, \bibinfo {author} {\bibfnamefont {A.}~\bibnamefont {Tsukazaki}},
  \bibinfo {author} {\bibfnamefont {R.}~\bibnamefont {Yoshimi}}, \bibinfo
  {author} {\bibfnamefont {K.~S.}\ \bibnamefont {Takahashi}}, \bibinfo {author}
  {\bibfnamefont {M.}~\bibnamefont {Kawasaki}},\ and\ \bibinfo {author}
  {\bibfnamefont {Y.}~\bibnamefont {Tokura}},\ }\bibfield  {title} {\bibinfo
  {title} {Large unidirectional magnetoresistance in a magnetic topological
  insulator},\ }\href {https://doi.org/10.1103/PhysRevLett.117.127202}
  {\bibfield  {journal} {\bibinfo  {journal} {Phys. Rev. Lett.}\ }\textbf
  {\bibinfo {volume} {117}},\ \bibinfo {pages} {127202} (\bibinfo {year}
  {2016})}\BibitemShut {NoStop}%
\bibitem [{\citenamefont {Krstić}\ \emph {et~al.}(2002)\citenamefont
  {Krstić}, \citenamefont {Roth}, \citenamefont {Burghard}, \citenamefont
  {Kern},\ and\ \citenamefont {Rikken}}]{Krstic2002}%
  \BibitemOpen
  \bibfield  {author} {\bibinfo {author} {\bibfnamefont {V.}~\bibnamefont
  {Krstić}}, \bibinfo {author} {\bibfnamefont {S.}~\bibnamefont {Roth}},
  \bibinfo {author} {\bibfnamefont {M.}~\bibnamefont {Burghard}}, \bibinfo
  {author} {\bibfnamefont {K.}~\bibnamefont {Kern}},\ and\ \bibinfo {author}
  {\bibfnamefont {G.~L. J.~A.}\ \bibnamefont {Rikken}},\ }\bibfield  {title}
  {\bibinfo {title} {Magneto-chiral anisotropy in charge transport through
  single-walled carbon nanotubes},\ }\href {https://doi.org/10.1063/1.1523895}
  {\bibfield  {journal} {\bibinfo  {journal} {The Journal of Chemical Physics}\
  }\textbf {\bibinfo {volume} {117}},\ \bibinfo {pages} {11315} (\bibinfo
  {year} {2002})},\ \Eprint
  {https://arxiv.org/abs/https://doi.org/10.1063/1.1523895}
  {https://doi.org/10.1063/1.1523895} \BibitemShut {NoStop}%
\bibitem [{\citenamefont {Pop}\ \emph {et~al.}(2014)\citenamefont {Pop},
  \citenamefont {Auban-Senzier}, \citenamefont {Canadell}, \citenamefont
  {Rikken},\ and\ \citenamefont {Avarvari}}]{Pop2014}%
  \BibitemOpen
  \bibfield  {author} {\bibinfo {author} {\bibfnamefont {F.}~\bibnamefont
  {Pop}}, \bibinfo {author} {\bibfnamefont {P.}~\bibnamefont {Auban-Senzier}},
  \bibinfo {author} {\bibfnamefont {E.}~\bibnamefont {Canadell}}, \bibinfo
  {author} {\bibfnamefont {G.~L. J.~A.}\ \bibnamefont {Rikken}},\ and\ \bibinfo
  {author} {\bibfnamefont {N.}~\bibnamefont {Avarvari}},\ }\bibfield  {title}
  {\bibinfo {title} {Electrical magnetochiral anisotropy in a bulk chiral
  molecular conductor},\ }\href {https://doi.org/10.1038/ncomms4757} {\bibfield
   {journal} {\bibinfo  {journal} {Nature Communications}\ }\textbf {\bibinfo
  {volume} {5}},\ \bibinfo {pages} {3757} (\bibinfo {year} {2014})}\BibitemShut
  {NoStop}%
\bibitem [{\citenamefont {Yokouchi}\ \emph {et~al.}(2017)\citenamefont
  {Yokouchi}, \citenamefont {Kanazawa}, \citenamefont {Kikkawa}, \citenamefont
  {Morikawa}, \citenamefont {Shibata}, \citenamefont {Arima}, \citenamefont
  {Taguchi}, \citenamefont {Kagawa},\ and\ \citenamefont
  {Tokura}}]{Yokouchi2017}%
  \BibitemOpen
  \bibfield  {author} {\bibinfo {author} {\bibfnamefont {T.}~\bibnamefont
  {Yokouchi}}, \bibinfo {author} {\bibfnamefont {N.}~\bibnamefont {Kanazawa}},
  \bibinfo {author} {\bibfnamefont {A.}~\bibnamefont {Kikkawa}}, \bibinfo
  {author} {\bibfnamefont {D.}~\bibnamefont {Morikawa}}, \bibinfo {author}
  {\bibfnamefont {K.}~\bibnamefont {Shibata}}, \bibinfo {author} {\bibfnamefont
  {T.}~\bibnamefont {Arima}}, \bibinfo {author} {\bibfnamefont
  {Y.}~\bibnamefont {Taguchi}}, \bibinfo {author} {\bibfnamefont
  {F.}~\bibnamefont {Kagawa}},\ and\ \bibinfo {author} {\bibfnamefont
  {Y.}~\bibnamefont {Tokura}},\ }\bibfield  {title} {\bibinfo {title}
  {Electrical magnetochiral effect induced by chiral spin fluctuations},\
  }\href {https://doi.org/10.1038/s41467-017-01094-2} {\bibfield  {journal}
  {\bibinfo  {journal} {Nature Communications}\ }\textbf {\bibinfo {volume}
  {8}},\ \bibinfo {pages} {866} (\bibinfo {year} {2017})}\BibitemShut {NoStop}%
\bibitem [{\citenamefont {Aoki}\ \emph {et~al.}(2019)\citenamefont {Aoki},
  \citenamefont {Kousaka},\ and\ \citenamefont {Togawa}}]{Aoki2019}%
  \BibitemOpen
  \bibfield  {author} {\bibinfo {author} {\bibfnamefont {R.}~\bibnamefont
  {Aoki}}, \bibinfo {author} {\bibfnamefont {Y.}~\bibnamefont {Kousaka}},\ and\
  \bibinfo {author} {\bibfnamefont {Y.}~\bibnamefont {Togawa}},\ }\bibfield
  {title} {\bibinfo {title} {Anomalous nonreciprocal electrical transport on
  chiral magnetic order},\ }\href
  {https://doi.org/10.1103/PhysRevLett.122.057206} {\bibfield  {journal}
  {\bibinfo  {journal} {Phys. Rev. Lett.}\ }\textbf {\bibinfo {volume} {122}},\
  \bibinfo {pages} {057206} (\bibinfo {year} {2019})}\BibitemShut {NoStop}%
\bibitem [{\citenamefont {Kitaori}\ \emph {et~al.}(2021)\citenamefont
  {Kitaori}, \citenamefont {Kanazawa}, \citenamefont {Ishizuka}, \citenamefont
  {Yokouchi}, \citenamefont {Nagaosa},\ and\ \citenamefont
  {Tokura}}]{Kitaori2021}%
  \BibitemOpen
  \bibfield  {author} {\bibinfo {author} {\bibfnamefont {A.}~\bibnamefont
  {Kitaori}}, \bibinfo {author} {\bibfnamefont {N.}~\bibnamefont {Kanazawa}},
  \bibinfo {author} {\bibfnamefont {H.}~\bibnamefont {Ishizuka}}, \bibinfo
  {author} {\bibfnamefont {T.}~\bibnamefont {Yokouchi}}, \bibinfo {author}
  {\bibfnamefont {N.}~\bibnamefont {Nagaosa}},\ and\ \bibinfo {author}
  {\bibfnamefont {Y.}~\bibnamefont {Tokura}},\ }\bibfield  {title} {\bibinfo
  {title} {Enhanced electrical magnetochiral effect by spin-hedgehog lattice
  structural transition},\ }\href
  {https://doi.org/10.1103/PhysRevB.103.L220410} {\bibfield  {journal}
  {\bibinfo  {journal} {Phys. Rev. B}\ }\textbf {\bibinfo {volume} {103}},\
  \bibinfo {pages} {L220410} (\bibinfo {year} {2021})}\BibitemShut {NoStop}%
\bibitem [{\citenamefont {Morimoto}\ and\ \citenamefont
  {Nagaosa}(2016{\natexlab{a}})}]{Morimoto20162}%
  \BibitemOpen
  \bibfield  {author} {\bibinfo {author} {\bibfnamefont {T.}~\bibnamefont
  {Morimoto}}\ and\ \bibinfo {author} {\bibfnamefont {N.}~\bibnamefont
  {Nagaosa}},\ }\bibfield  {title} {\bibinfo {title} {Chiral anomaly and giant
  magnetochiral anisotropy in noncentrosymmetric weyl semimetals},\ }\href
  {https://doi.org/10.1103/PhysRevLett.117.146603} {\bibfield  {journal}
  {\bibinfo  {journal} {Phys. Rev. Lett.}\ }\textbf {\bibinfo {volume} {117}},\
  \bibinfo {pages} {146603} (\bibinfo {year} {2016}{\natexlab{a}})}\BibitemShut
  {NoStop}%
\bibitem [{\citenamefont {Wakatsuki}\ \emph {et~al.}(2017)\citenamefont
  {Wakatsuki}, \citenamefont {Saito}, \citenamefont {Hoshino}, \citenamefont
  {Itahashi}, \citenamefont {Ideue}, \citenamefont {Ezawa}, \citenamefont
  {Iwasa},\ and\ \citenamefont {Nagaosa}}]{Wakatsukie1602390}%
  \BibitemOpen
  \bibfield  {author} {\bibinfo {author} {\bibfnamefont {R.}~\bibnamefont
  {Wakatsuki}}, \bibinfo {author} {\bibfnamefont {Y.}~\bibnamefont {Saito}},
  \bibinfo {author} {\bibfnamefont {S.}~\bibnamefont {Hoshino}}, \bibinfo
  {author} {\bibfnamefont {Y.~M.}\ \bibnamefont {Itahashi}}, \bibinfo {author}
  {\bibfnamefont {T.}~\bibnamefont {Ideue}}, \bibinfo {author} {\bibfnamefont
  {M.}~\bibnamefont {Ezawa}}, \bibinfo {author} {\bibfnamefont
  {Y.}~\bibnamefont {Iwasa}},\ and\ \bibinfo {author} {\bibfnamefont
  {N.}~\bibnamefont {Nagaosa}},\ }\bibfield  {title} {\bibinfo {title}
  {Nonreciprocal charge transport in noncentrosymmetric superconductors},\
  }\bibfield  {journal} {\bibinfo  {journal} {Science Advances}\ }\textbf
  {\bibinfo {volume} {3}},\ \href {https://doi.org/10.1126/sciadv.1602390}
  {10.1126/sciadv.1602390} (\bibinfo {year} {2017}),\ \Eprint
  {https://arxiv.org/abs/https://advances.sciencemag.org/content/3/4/e1602390.full.pdf}
  {https://advances.sciencemag.org/content/3/4/e1602390.full.pdf} \BibitemShut
  {NoStop}%
\bibitem [{\citenamefont {Ishizuka}\ and\ \citenamefont
  {Nagaosa}(2020)}]{Ishizuka2020}%
  \BibitemOpen
  \bibfield  {author} {\bibinfo {author} {\bibfnamefont {H.}~\bibnamefont
  {Ishizuka}}\ and\ \bibinfo {author} {\bibfnamefont {N.}~\bibnamefont
  {Nagaosa}},\ }\bibfield  {title} {\bibinfo {title} {Anomalous electrical
  magnetochiral effect by chiral spin-cluster scattering},\ }\href
  {https://doi.org/10.1038/s41467-020-16751-2} {\bibfield  {journal} {\bibinfo
  {journal} {Nature Communications}\ }\textbf {\bibinfo {volume} {11}},\
  \bibinfo {pages} {2986} (\bibinfo {year} {2020})}\BibitemShut {NoStop}%
\bibitem [{\citenamefont {Sipe}\ and\ \citenamefont
  {Shkrebtii}(2000)}]{Sipe2000}%
  \BibitemOpen
  \bibfield  {author} {\bibinfo {author} {\bibfnamefont {J.~E.}\ \bibnamefont
  {Sipe}}\ and\ \bibinfo {author} {\bibfnamefont {A.~I.}\ \bibnamefont
  {Shkrebtii}},\ }\bibfield  {title} {\bibinfo {title} {Second-order optical
  response in semiconductors},\ }\href
  {https://doi.org/10.1103/PhysRevB.61.5337} {\bibfield  {journal} {\bibinfo
  {journal} {Phys. Rev. B}\ }\textbf {\bibinfo {volume} {61}},\ \bibinfo
  {pages} {5337} (\bibinfo {year} {2000})}\BibitemShut {NoStop}%
\bibitem [{\citenamefont {Morimoto}\ and\ \citenamefont
  {Nagaosa}(2016{\natexlab{b}})}]{Morimoto2016}%
  \BibitemOpen
  \bibfield  {author} {\bibinfo {author} {\bibfnamefont {T.}~\bibnamefont
  {Morimoto}}\ and\ \bibinfo {author} {\bibfnamefont {N.}~\bibnamefont
  {Nagaosa}},\ }\bibfield  {title} {\bibinfo {title} {Topological nature of
  nonlinear optical effects in solids},\ }\href
  {https://advances.sciencemag.org/content/2/5/e1501524} {\bibfield  {journal}
  {\bibinfo  {journal} {Science Advances}\ }\textbf {\bibinfo {volume} {2}}
  (\bibinfo {year} {2016}{\natexlab{b}})}\BibitemShut {NoStop}%
\bibitem [{\citenamefont {Nagaosa}\ and\ \citenamefont
  {Morimoto}(2017)}]{Nagaosa2017}%
  \BibitemOpen
  \bibfield  {author} {\bibinfo {author} {\bibfnamefont {N.}~\bibnamefont
  {Nagaosa}}\ and\ \bibinfo {author} {\bibfnamefont {T.}~\bibnamefont
  {Morimoto}},\ }\bibfield  {title} {\bibinfo {title} {Concept of quantum
  geometry in optoelectronic processes in solids: Application to solar cells},\
  }\href {https://doi.org/https://doi.org/10.1002/adma.201603345} {\bibfield
  {journal} {\bibinfo  {journal} {Advanced Materials}\ }\textbf {\bibinfo
  {volume} {29}},\ \bibinfo {pages} {1603345} (\bibinfo {year}
  {2017})}\BibitemShut {NoStop}%
\bibitem [{\citenamefont {Wu}\ \emph {et~al.}(2017)\citenamefont {Wu},
  \citenamefont {Patankar}, \citenamefont {Morimoto}, \citenamefont {Nair},
  \citenamefont {Thewalt}, \citenamefont {Little}, \citenamefont {Analytis},
  \citenamefont {Moore},\ and\ \citenamefont {Orenstein}}]{Wu2017}%
  \BibitemOpen
  \bibfield  {author} {\bibinfo {author} {\bibfnamefont {L.}~\bibnamefont
  {Wu}}, \bibinfo {author} {\bibfnamefont {S.}~\bibnamefont {Patankar}},
  \bibinfo {author} {\bibfnamefont {T.}~\bibnamefont {Morimoto}}, \bibinfo
  {author} {\bibfnamefont {N.~L.}\ \bibnamefont {Nair}}, \bibinfo {author}
  {\bibfnamefont {E.}~\bibnamefont {Thewalt}}, \bibinfo {author} {\bibfnamefont
  {A.}~\bibnamefont {Little}}, \bibinfo {author} {\bibfnamefont {J.~G.}\
  \bibnamefont {Analytis}}, \bibinfo {author} {\bibfnamefont {J.~E.}\
  \bibnamefont {Moore}},\ and\ \bibinfo {author} {\bibfnamefont
  {J.}~\bibnamefont {Orenstein}},\ }\bibfield  {title} {\bibinfo {title} {Giant
  anisotropic nonlinear optical response in transition metal monopnictide weyl
  semimetals},\ }\href {https://doi.org/10.1038/nphys3969} {\bibfield
  {journal} {\bibinfo  {journal} {Nature Physics}\ }\textbf {\bibinfo {volume}
  {13}},\ \bibinfo {pages} {350} (\bibinfo {year} {2017})}\BibitemShut
  {NoStop}%
\bibitem [{\citenamefont {Moore}\ and\ \citenamefont
  {Orenstein}(2010)}]{Moore2010}%
  \BibitemOpen
  \bibfield  {author} {\bibinfo {author} {\bibfnamefont {J.~E.}\ \bibnamefont
  {Moore}}\ and\ \bibinfo {author} {\bibfnamefont {J.}~\bibnamefont
  {Orenstein}},\ }\bibfield  {title} {\bibinfo {title} {Confinement-induced
  berry phase and helicity-dependent photocurrents},\ }\href
  {https://doi.org/10.1103/PhysRevLett.105.026805} {\bibfield  {journal}
  {\bibinfo  {journal} {Phys. Rev. Lett.}\ }\textbf {\bibinfo {volume} {105}},\
  \bibinfo {pages} {026805} (\bibinfo {year} {2010})}\BibitemShut {NoStop}%
\bibitem [{\citenamefont {Sodemann}\ and\ \citenamefont
  {Fu}(2015)}]{Sodemann2015}%
  \BibitemOpen
  \bibfield  {author} {\bibinfo {author} {\bibfnamefont {I.}~\bibnamefont
  {Sodemann}}\ and\ \bibinfo {author} {\bibfnamefont {L.}~\bibnamefont {Fu}},\
  }\bibfield  {title} {\bibinfo {title} {Quantum nonlinear hall effect induced
  by berry curvature dipole in time-reversal invariant materials},\ }\href
  {https://doi.org/10.1103/PhysRevLett.115.216806} {\bibfield  {journal}
  {\bibinfo  {journal} {Phys. Rev. Lett.}\ }\textbf {\bibinfo {volume} {115}},\
  \bibinfo {pages} {216806} (\bibinfo {year} {2015})}\BibitemShut {NoStop}%
\bibitem [{\citenamefont {Tokura}\ and\ \citenamefont
  {Nagaosa}(2018)}]{Tokura2018}%
  \BibitemOpen
  \bibfield  {author} {\bibinfo {author} {\bibfnamefont {Y.}~\bibnamefont
  {Tokura}}\ and\ \bibinfo {author} {\bibfnamefont {N.}~\bibnamefont
  {Nagaosa}},\ }\bibfield  {title} {\bibinfo {title} {Nonreciprocal responses
  from non-centrosymmetric quantum materials},\ }\href
  {https://doi.org/10.1038/s41467-018-05759-4} {\bibfield  {journal} {\bibinfo
  {journal} {Nature Communications}\ }\textbf {\bibinfo {volume} {9}},\
  \bibinfo {pages} {3740} (\bibinfo {year} {2018})}\BibitemShut {NoStop}%
\bibitem [{\citenamefont {Dzsaber}\ \emph {et~al.}(2021)\citenamefont
  {Dzsaber}, \citenamefont {Yan}, \citenamefont {Taupin}, \citenamefont
  {Eguchi}, \citenamefont {Prokofiev}, \citenamefont {Shiroka}, \citenamefont
  {Blaha}, \citenamefont {Rubel}, \citenamefont {Grefe}, \citenamefont {Lai},
  \citenamefont {Si},\ and\ \citenamefont {Paschen}}]{Dzsaber2021}%
  \BibitemOpen
  \bibfield  {author} {\bibinfo {author} {\bibfnamefont {S.}~\bibnamefont
  {Dzsaber}}, \bibinfo {author} {\bibfnamefont {X.}~\bibnamefont {Yan}},
  \bibinfo {author} {\bibfnamefont {M.}~\bibnamefont {Taupin}}, \bibinfo
  {author} {\bibfnamefont {G.}~\bibnamefont {Eguchi}}, \bibinfo {author}
  {\bibfnamefont {A.}~\bibnamefont {Prokofiev}}, \bibinfo {author}
  {\bibfnamefont {T.}~\bibnamefont {Shiroka}}, \bibinfo {author} {\bibfnamefont
  {P.}~\bibnamefont {Blaha}}, \bibinfo {author} {\bibfnamefont
  {O.}~\bibnamefont {Rubel}}, \bibinfo {author} {\bibfnamefont {S.~E.}\
  \bibnamefont {Grefe}}, \bibinfo {author} {\bibfnamefont {H.-H.}\ \bibnamefont
  {Lai}}, \bibinfo {author} {\bibfnamefont {Q.}~\bibnamefont {Si}},\ and\
  \bibinfo {author} {\bibfnamefont {S.}~\bibnamefont {Paschen}},\ }\bibfield
  {title} {\bibinfo {title} {Giant spontaneous hall effect in a nonmagnetic
  weyl{\textendash}kondo semimetal},\ }\bibfield  {journal} {\bibinfo
  {journal} {Proceedings of the National Academy of Sciences}\ }\textbf
  {\bibinfo {volume} {118}},\ \href {https://doi.org/10.1073/pnas.2013386118}
  {10.1073/pnas.2013386118} (\bibinfo {year} {2021}),\ \Eprint
  {https://arxiv.org/abs/https://www.pnas.org/content/118/8/e2013386118.full.pdf}
  {https://www.pnas.org/content/118/8/e2013386118.full.pdf} \BibitemShut
  {NoStop}%
\bibitem [{\citenamefont {Michishita}\ and\ \citenamefont
  {Peters}(2021)}]{Michishita2021}%
  \BibitemOpen
  \bibfield  {author} {\bibinfo {author} {\bibfnamefont {Y.}~\bibnamefont
  {Michishita}}\ and\ \bibinfo {author} {\bibfnamefont {R.}~\bibnamefont
  {Peters}},\ }\bibfield  {title} {\bibinfo {title} {Effects of renormalization
  and non-hermiticity on nonlinear responses in strongly correlated electron
  systems},\ }\href {https://doi.org/10.1103/PhysRevB.103.195133} {\bibfield
  {journal} {\bibinfo  {journal} {Phys. Rev. B}\ }\textbf {\bibinfo {volume}
  {103}},\ \bibinfo {pages} {195133} (\bibinfo {year} {2021})}\BibitemShut
  {NoStop}%
\bibitem [{\citenamefont {Kofuji}\ \emph {et~al.}(2021)\citenamefont {Kofuji},
  \citenamefont {Michishita},\ and\ \citenamefont {Peters}}]{Kofuji2021}%
  \BibitemOpen
  \bibfield  {author} {\bibinfo {author} {\bibfnamefont {A.}~\bibnamefont
  {Kofuji}}, \bibinfo {author} {\bibfnamefont {Y.}~\bibnamefont {Michishita}},\
  and\ \bibinfo {author} {\bibfnamefont {R.}~\bibnamefont {Peters}},\
  }\bibfield  {title} {\bibinfo {title} {Effects of strong correlations on the
  nonlinear response in weyl-kondo semimetals},\ }\href
  {https://doi.org/10.1103/PhysRevB.104.085151} {\bibfield  {journal} {\bibinfo
   {journal} {Phys. Rev. B}\ }\textbf {\bibinfo {volume} {104}},\ \bibinfo
  {pages} {085151} (\bibinfo {year} {2021})}\BibitemShut {NoStop}%
\bibitem [{\citenamefont {Metzner}\ and\ \citenamefont
  {Vollhardt}(1989)}]{Metzner1989}%
  \BibitemOpen
  \bibfield  {author} {\bibinfo {author} {\bibfnamefont {W.}~\bibnamefont
  {Metzner}}\ and\ \bibinfo {author} {\bibfnamefont {D.}~\bibnamefont
  {Vollhardt}},\ }\bibfield  {title} {\bibinfo {title} {Correlated lattice
  fermions in $d=\ensuremath{\infty}$ dimensions},\ }\href
  {https://doi.org/10.1103/PhysRevLett.62.324} {\bibfield  {journal} {\bibinfo
  {journal} {Phys. Rev. Lett.}\ }\textbf {\bibinfo {volume} {62}},\ \bibinfo
  {pages} {324} (\bibinfo {year} {1989})}\BibitemShut {NoStop}%
\bibitem [{\citenamefont {Georges}\ \emph {et~al.}(1996)\citenamefont
  {Georges}, \citenamefont {Kotliar}, \citenamefont {Krauth},\ and\
  \citenamefont {Rozenberg}}]{Georges1996}%
  \BibitemOpen
  \bibfield  {author} {\bibinfo {author} {\bibfnamefont {A.}~\bibnamefont
  {Georges}}, \bibinfo {author} {\bibfnamefont {G.}~\bibnamefont {Kotliar}},
  \bibinfo {author} {\bibfnamefont {W.}~\bibnamefont {Krauth}},\ and\ \bibinfo
  {author} {\bibfnamefont {M.~J.}\ \bibnamefont {Rozenberg}},\ }\bibfield
  {title} {\bibinfo {title} {Dynamical mean-field theory of strongly correlated
  fermion systems and the limit of infinite dimensions},\ }\href
  {https://doi.org/10.1103/RevModPhys.68.13} {\bibfield  {journal} {\bibinfo
  {journal} {Rev. Mod. Phys.}\ }\textbf {\bibinfo {volume} {68}},\ \bibinfo
  {pages} {13} (\bibinfo {year} {1996})}\BibitemShut {NoStop}%
\bibitem [{\citenamefont {Pruschke}\ \emph {et~al.}(1995)\citenamefont
  {Pruschke}, \citenamefont {Jarrell},\ and\ \citenamefont
  {Freericks}}]{Pruschke1995}%
  \BibitemOpen
  \bibfield  {author} {\bibinfo {author} {\bibfnamefont {T.}~\bibnamefont
  {Pruschke}}, \bibinfo {author} {\bibfnamefont {M.}~\bibnamefont {Jarrell}},\
  and\ \bibinfo {author} {\bibfnamefont {J.}~\bibnamefont {Freericks}},\
  }\bibfield  {title} {\bibinfo {title} {Anomalous normal-state properties of
  high-t c superconductors: intrinsic properties of strongly correlated
  electron systems?},\ }\href {https://doi.org/10.1080/00018739500101526}
  {\bibfield  {journal} {\bibinfo  {journal} {Advances in Physics}\ }\textbf
  {\bibinfo {volume} {44}},\ \bibinfo {pages} {187} (\bibinfo {year} {1995})},\
  \Eprint {https://arxiv.org/abs/https://doi.org/10.1080/00018739500101526}
  {https://doi.org/10.1080/00018739500101526} \BibitemShut {NoStop}%
\bibitem [{\citenamefont {Sugitani}\ \emph {et~al.}(2006)\citenamefont
  {Sugitani}, \citenamefont {Okuda}, \citenamefont {Shishido}, \citenamefont
  {Yamada}, \citenamefont {Thamizhavel}, \citenamefont {Yamamoto},
  \citenamefont {D.~Matsuda}, \citenamefont {Haga}, \citenamefont {Takeuchi},
  \citenamefont {Settai},\ and\ \citenamefont {Ōnuki}}]{JPSJ.75.043703}%
  \BibitemOpen
  \bibfield  {author} {\bibinfo {author} {\bibfnamefont {I.}~\bibnamefont
  {Sugitani}}, \bibinfo {author} {\bibfnamefont {Y.}~\bibnamefont {Okuda}},
  \bibinfo {author} {\bibfnamefont {H.}~\bibnamefont {Shishido}}, \bibinfo
  {author} {\bibfnamefont {T.}~\bibnamefont {Yamada}}, \bibinfo {author}
  {\bibfnamefont {A.}~\bibnamefont {Thamizhavel}}, \bibinfo {author}
  {\bibfnamefont {E.}~\bibnamefont {Yamamoto}}, \bibinfo {author}
  {\bibfnamefont {T.}~\bibnamefont {D.~Matsuda}}, \bibinfo {author}
  {\bibfnamefont {Y.}~\bibnamefont {Haga}}, \bibinfo {author} {\bibfnamefont
  {T.}~\bibnamefont {Takeuchi}}, \bibinfo {author} {\bibfnamefont
  {R.}~\bibnamefont {Settai}},\ and\ \bibinfo {author} {\bibfnamefont
  {Y.}~\bibnamefont {Ōnuki}},\ }\bibfield  {title} {\bibinfo {title}
  {Pressure-induced heavy-fermion superconductivity in antiferromagnet ceirsi3
  without inversion symmetry},\ }\href {https://doi.org/10.1143/JPSJ.75.043703}
  {\bibfield  {journal} {\bibinfo  {journal} {Journal of the Physical Society
  of Japan}\ }\textbf {\bibinfo {volume} {75}},\ \bibinfo {pages} {043703}
  (\bibinfo {year} {2006})},\ \Eprint
  {https://arxiv.org/abs/https://doi.org/10.1143/JPSJ.75.043703}
  {https://doi.org/10.1143/JPSJ.75.043703} \BibitemShut {NoStop}%
\bibitem [{\citenamefont {Kimura}\ \emph {et~al.}(2005)\citenamefont {Kimura},
  \citenamefont {Ito}, \citenamefont {Saitoh}, \citenamefont {Umeda},
  \citenamefont {Aoki},\ and\ \citenamefont
  {Terashima}}]{PhysRevLett.95.247004}%
  \BibitemOpen
  \bibfield  {author} {\bibinfo {author} {\bibfnamefont {N.}~\bibnamefont
  {Kimura}}, \bibinfo {author} {\bibfnamefont {K.}~\bibnamefont {Ito}},
  \bibinfo {author} {\bibfnamefont {K.}~\bibnamefont {Saitoh}}, \bibinfo
  {author} {\bibfnamefont {Y.}~\bibnamefont {Umeda}}, \bibinfo {author}
  {\bibfnamefont {H.}~\bibnamefont {Aoki}},\ and\ \bibinfo {author}
  {\bibfnamefont {T.}~\bibnamefont {Terashima}},\ }\bibfield  {title} {\bibinfo
  {title} {Pressure-induced superconductivity in noncentrosymmetric
  heavy-fermion ${\mathrm{cerhsi}}_{3}$},\ }\href
  {https://doi.org/10.1103/PhysRevLett.95.247004} {\bibfield  {journal}
  {\bibinfo  {journal} {Phys. Rev. Lett.}\ }\textbf {\bibinfo {volume} {95}},\
  \bibinfo {pages} {247004} (\bibinfo {year} {2005})}\BibitemShut {NoStop}%
\bibitem [{\citenamefont {Bauer}\ \emph {et~al.}(2004)\citenamefont {Bauer},
  \citenamefont {Hilscher}, \citenamefont {Michor}, \citenamefont {Paul},
  \citenamefont {Scheidt}, \citenamefont {Gribanov}, \citenamefont {Seropegin},
  \citenamefont {No\"el}, \citenamefont {Sigrist},\ and\ \citenamefont
  {Rogl}}]{PhysRevLett.92.027003}%
  \BibitemOpen
  \bibfield  {author} {\bibinfo {author} {\bibfnamefont {E.}~\bibnamefont
  {Bauer}}, \bibinfo {author} {\bibfnamefont {G.}~\bibnamefont {Hilscher}},
  \bibinfo {author} {\bibfnamefont {H.}~\bibnamefont {Michor}}, \bibinfo
  {author} {\bibfnamefont {C.}~\bibnamefont {Paul}}, \bibinfo {author}
  {\bibfnamefont {E.~W.}\ \bibnamefont {Scheidt}}, \bibinfo {author}
  {\bibfnamefont {A.}~\bibnamefont {Gribanov}}, \bibinfo {author}
  {\bibfnamefont {Y.}~\bibnamefont {Seropegin}}, \bibinfo {author}
  {\bibfnamefont {H.}~\bibnamefont {No\"el}}, \bibinfo {author} {\bibfnamefont
  {M.}~\bibnamefont {Sigrist}},\ and\ \bibinfo {author} {\bibfnamefont
  {P.}~\bibnamefont {Rogl}},\ }\bibfield  {title} {\bibinfo {title} {Heavy
  fermion superconductivity and magnetic order in noncentrosymmetric
  ${\mathrm{c}\mathrm{e}\mathrm{p}\mathrm{t}}_{3}\mathrm{S}\mathrm{i}$},\
  }\href {https://doi.org/10.1103/PhysRevLett.92.027003} {\bibfield  {journal}
  {\bibinfo  {journal} {Phys. Rev. Lett.}\ }\textbf {\bibinfo {volume} {92}},\
  \bibinfo {pages} {027003} (\bibinfo {year} {2004})}\BibitemShut {NoStop}%
\bibitem [{\citenamefont {Maruyama}\ and\ \citenamefont
  {Yanase}(2015)}]{JPSJ.84.074702}%
  \BibitemOpen
  \bibfield  {author} {\bibinfo {author} {\bibfnamefont {D.}~\bibnamefont
  {Maruyama}}\ and\ \bibinfo {author} {\bibfnamefont {Y.}~\bibnamefont
  {Yanase}},\ }\bibfield  {title} {\bibinfo {title} {Electron correlation
  effects in non-centrosymmetric metals in the weak coupling regime},\ }\href
  {https://doi.org/10.7566/JPSJ.84.074702} {\bibfield  {journal} {\bibinfo
  {journal} {Journal of the Physical Society of Japan}\ }\textbf {\bibinfo
  {volume} {84}},\ \bibinfo {pages} {074702} (\bibinfo {year} {2015})},\
  \Eprint {https://arxiv.org/abs/https://doi.org/10.7566/JPSJ.84.074702}
  {https://doi.org/10.7566/JPSJ.84.074702} \BibitemShut {NoStop}%
\bibitem [{\citenamefont {Michishita}\ and\ \citenamefont
  {Peters}(2019)}]{PhysRevB.99.155141}%
  \BibitemOpen
  \bibfield  {author} {\bibinfo {author} {\bibfnamefont {Y.}~\bibnamefont
  {Michishita}}\ and\ \bibinfo {author} {\bibfnamefont {R.}~\bibnamefont
  {Peters}},\ }\bibfield  {title} {\bibinfo {title} {Impact of the rashba
  spin-orbit coupling on $f$-electron materials},\ }\href
  {https://doi.org/10.1103/PhysRevB.99.155141} {\bibfield  {journal} {\bibinfo
  {journal} {Phys. Rev. B}\ }\textbf {\bibinfo {volume} {99}},\ \bibinfo
  {pages} {155141} (\bibinfo {year} {2019})}\BibitemShut {NoStop}%
\bibitem [{\citenamefont {Schrieffer}\ and\ \citenamefont
  {Wolff}(1966)}]{Schrieffer1966}%
  \BibitemOpen
  \bibfield  {author} {\bibinfo {author} {\bibfnamefont {J.~R.}\ \bibnamefont
  {Schrieffer}}\ and\ \bibinfo {author} {\bibfnamefont {P.~A.}\ \bibnamefont
  {Wolff}},\ }\bibfield  {title} {\bibinfo {title} {Relation between the
  anderson and kondo hamiltonians},\ }\href
  {https://doi.org/10.1103/PhysRev.149.491} {\bibfield  {journal} {\bibinfo
  {journal} {Phys. Rev.}\ }\textbf {\bibinfo {volume} {149}},\ \bibinfo {pages}
  {491} (\bibinfo {year} {1966})}\BibitemShut {NoStop}%
\bibitem [{\citenamefont {Isobe}\ \emph {et~al.}(2020)\citenamefont {Isobe},
  \citenamefont {Xu},\ and\ \citenamefont {Fu}}]{Isobe2020}%
  \BibitemOpen
  \bibfield  {author} {\bibinfo {author} {\bibfnamefont {H.}~\bibnamefont
  {Isobe}}, \bibinfo {author} {\bibfnamefont {S.-Y.}\ \bibnamefont {Xu}},\ and\
  \bibinfo {author} {\bibfnamefont {L.}~\bibnamefont {Fu}},\ }\bibfield
  {title} {\bibinfo {title} {High-frequency rectification via chiral bloch
  electrons},\ }\href {https://doi.org/10.1126/sciadv.aay2497} {\bibfield
  {journal} {\bibinfo  {journal} {Science Advances}\ }\textbf {\bibinfo
  {volume} {6}},\ \bibinfo {pages} {eaay2497} (\bibinfo {year} {2020})},\
  \Eprint
  {https://arxiv.org/abs/https://www.science.org/doi/pdf/10.1126/sciadv.aay2497}
  {https://www.science.org/doi/pdf/10.1126/sciadv.aay2497} \BibitemShut
  {NoStop}%
\bibitem [{\citenamefont {Wilson}(1975)}]{Wilson1975}%
  \BibitemOpen
  \bibfield  {author} {\bibinfo {author} {\bibfnamefont {K.~G.}\ \bibnamefont
  {Wilson}},\ }\bibfield  {title} {\bibinfo {title} {The renormalization group:
  Critical phenomena and the kondo problem},\ }\href
  {https://doi.org/10.1103/RevModPhys.47.773} {\bibfield  {journal} {\bibinfo
  {journal} {Rev. Mod. Phys.}\ }\textbf {\bibinfo {volume} {47}},\ \bibinfo
  {pages} {773} (\bibinfo {year} {1975})}\BibitemShut {NoStop}%
\bibitem [{\citenamefont {Bulla}\ \emph {et~al.}(2008)\citenamefont {Bulla},
  \citenamefont {Costi},\ and\ \citenamefont {Pruschke}}]{Bulla2008}%
  \BibitemOpen
  \bibfield  {author} {\bibinfo {author} {\bibfnamefont {R.}~\bibnamefont
  {Bulla}}, \bibinfo {author} {\bibfnamefont {T.~A.}\ \bibnamefont {Costi}},\
  and\ \bibinfo {author} {\bibfnamefont {T.}~\bibnamefont {Pruschke}},\
  }\bibfield  {title} {\bibinfo {title} {Numerical renormalization group method
  for quantum impurity systems},\ }\href
  {https://doi.org/10.1103/RevModPhys.80.395} {\bibfield  {journal} {\bibinfo
  {journal} {Rev. Mod. Phys.}\ }\textbf {\bibinfo {volume} {80}},\ \bibinfo
  {pages} {395} (\bibinfo {year} {2008})}\BibitemShut {NoStop}%
\bibitem [{\citenamefont {Peters}\ \emph {et~al.}(2006)\citenamefont {Peters},
  \citenamefont {Pruschke},\ and\ \citenamefont {Anders}}]{Peters2006}%
  \BibitemOpen
  \bibfield  {author} {\bibinfo {author} {\bibfnamefont {R.}~\bibnamefont
  {Peters}}, \bibinfo {author} {\bibfnamefont {T.}~\bibnamefont {Pruschke}},\
  and\ \bibinfo {author} {\bibfnamefont {F.~B.}\ \bibnamefont {Anders}},\
  }\bibfield  {title} {\bibinfo {title} {Numerical renormalization group
  approach to green's functions for quantum impurity models},\ }\href
  {https://doi.org/10.1103/PhysRevB.74.245114} {\bibfield  {journal} {\bibinfo
  {journal} {Phys. Rev. B}\ }\textbf {\bibinfo {volume} {74}},\ \bibinfo
  {pages} {245114} (\bibinfo {year} {2006})}\BibitemShut {NoStop}%
\bibitem [{\citenamefont {Weichselbaum}\ and\ \citenamefont {von
  Delft}(2007)}]{Weichselbaum2007}%
  \BibitemOpen
  \bibfield  {author} {\bibinfo {author} {\bibfnamefont {A.}~\bibnamefont
  {Weichselbaum}}\ and\ \bibinfo {author} {\bibfnamefont {J.}~\bibnamefont {von
  Delft}},\ }\bibfield  {title} {\bibinfo {title} {Sum-rule conserving spectral
  functions from the numerical renormalization group},\ }\href
  {https://doi.org/10.1103/PhysRevLett.99.076402} {\bibfield  {journal}
  {\bibinfo  {journal} {Phys. Rev. Lett.}\ }\textbf {\bibinfo {volume} {99}},\
  \bibinfo {pages} {076402} (\bibinfo {year} {2007})}\BibitemShut {NoStop}%
\bibitem [{\citenamefont {Doniach}(1977)}]{Doniach1977}%
  \BibitemOpen
  \bibfield  {author} {\bibinfo {author} {\bibfnamefont {S.}~\bibnamefont
  {Doniach}},\ }\bibfield  {title} {\bibinfo {title} {The kondo lattice and
  weak antiferromagnetism},\ }\href
  {https://doi.org/https://doi.org/10.1016/0378-4363(77)90190-5} {\bibfield
  {journal} {\bibinfo  {journal} {Physica B+C}\ }\textbf {\bibinfo {volume}
  {91}},\ \bibinfo {pages} {231} (\bibinfo {year} {1977})}\BibitemShut
  {NoStop}%
\bibitem [{\citenamefont {Tsunetsugu}\ \emph {et~al.}(1997)\citenamefont
  {Tsunetsugu}, \citenamefont {Sigrist},\ and\ \citenamefont
  {Ueda}}]{Tsunetsugu1997}%
  \BibitemOpen
  \bibfield  {author} {\bibinfo {author} {\bibfnamefont {H.}~\bibnamefont
  {Tsunetsugu}}, \bibinfo {author} {\bibfnamefont {M.}~\bibnamefont
  {Sigrist}},\ and\ \bibinfo {author} {\bibfnamefont {K.}~\bibnamefont
  {Ueda}},\ }\bibfield  {title} {\bibinfo {title} {The ground-state phase
  diagram of the one-dimensional kondo lattice model},\ }\href
  {https://doi.org/10.1103/RevModPhys.69.809} {\bibfield  {journal} {\bibinfo
  {journal} {Rev. Mod. Phys.}\ }\textbf {\bibinfo {volume} {69}},\ \bibinfo
  {pages} {809} (\bibinfo {year} {1997})}\BibitemShut {NoStop}%
\bibitem [{\citenamefont {Wagner}\ \emph {et~al.}(2020)\citenamefont {Wagner},
  \citenamefont {Wr{\'o}bel},\ and\ \citenamefont {Eder}}]{Wagner2020}%
  \BibitemOpen
  \bibfield  {author} {\bibinfo {author} {\bibfnamefont {P.}~\bibnamefont
  {Wagner}}, \bibinfo {author} {\bibfnamefont {P.}~\bibnamefont {Wr{\'o}bel}},\
  and\ \bibinfo {author} {\bibfnamefont {R.}~\bibnamefont {Eder}},\ }\bibfield
  {title} {\bibinfo {title} {Ferromagnetism in the kondo-lattice},\ }\href
  {https://doi.org/10.1140/epjb/e2020-100549-0} {\bibfield  {journal} {\bibinfo
   {journal} {The European Physical Journal B}\ }\textbf {\bibinfo {volume}
  {93}},\ \bibinfo {pages} {58} (\bibinfo {year} {2020})}\BibitemShut {NoStop}%
\bibitem [{\citenamefont {Bernhard}\ and\ \citenamefont
  {Lacroix}(2015)}]{PhysRevB.92.094401}%
  \BibitemOpen
  \bibfield  {author} {\bibinfo {author} {\bibfnamefont {B.~H.}\ \bibnamefont
  {Bernhard}}\ and\ \bibinfo {author} {\bibfnamefont {C.}~\bibnamefont
  {Lacroix}},\ }\bibfield  {title} {\bibinfo {title} {Coexistence of magnetic
  order and kondo effect in the kondo-heisenberg model},\ }\href
  {https://doi.org/10.1103/PhysRevB.92.094401} {\bibfield  {journal} {\bibinfo
  {journal} {Phys. Rev. B}\ }\textbf {\bibinfo {volume} {92}},\ \bibinfo
  {pages} {094401} (\bibinfo {year} {2015})}\BibitemShut {NoStop}%
\bibitem [{\citenamefont {Smerat}\ \emph {et~al.}(2009)\citenamefont {Smerat},
  \citenamefont {Schollw\"ock}, \citenamefont {McCulloch},\ and\ \citenamefont
  {Schoeller}}]{Smerat2009}%
  \BibitemOpen
  \bibfield  {author} {\bibinfo {author} {\bibfnamefont {S.}~\bibnamefont
  {Smerat}}, \bibinfo {author} {\bibfnamefont {U.}~\bibnamefont
  {Schollw\"ock}}, \bibinfo {author} {\bibfnamefont {I.~P.}\ \bibnamefont
  {McCulloch}},\ and\ \bibinfo {author} {\bibfnamefont {H.}~\bibnamefont
  {Schoeller}},\ }\bibfield  {title} {\bibinfo {title} {Quasiparticles in the
  kondo lattice model at partial fillings of the conduction band using the
  density matrix renormalization group},\ }\href
  {https://doi.org/10.1103/PhysRevB.79.235107} {\bibfield  {journal} {\bibinfo
  {journal} {Phys. Rev. B}\ }\textbf {\bibinfo {volume} {79}},\ \bibinfo
  {pages} {235107} (\bibinfo {year} {2009})}\BibitemShut {NoStop}%
\bibitem [{\citenamefont {Kubo}(2015)}]{JPSJ.84.094702}%
  \BibitemOpen
  \bibfield  {author} {\bibinfo {author} {\bibfnamefont {K.}~\bibnamefont
  {Kubo}},\ }\bibfield  {title} {\bibinfo {title} {Lifshitz transitions in
  magnetic phases of the periodic anderson model},\ }\href
  {https://doi.org/10.7566/JPSJ.84.094702} {\bibfield  {journal} {\bibinfo
  {journal} {Journal of the Physical Society of Japan}\ }\textbf {\bibinfo
  {volume} {84}},\ \bibinfo {pages} {094702} (\bibinfo {year} {2015})},\
  \Eprint {https://arxiv.org/abs/https://doi.org/10.7566/JPSJ.84.094702}
  {https://doi.org/10.7566/JPSJ.84.094702} \BibitemShut {NoStop}%
\bibitem [{\citenamefont {Li}\ \emph {et~al.}(2010)\citenamefont {Li},
  \citenamefont {Zhang},\ and\ \citenamefont {Yu}}]{Li2010}%
  \BibitemOpen
  \bibfield  {author} {\bibinfo {author} {\bibfnamefont {G.-B.}\ \bibnamefont
  {Li}}, \bibinfo {author} {\bibfnamefont {G.-M.}\ \bibnamefont {Zhang}},\ and\
  \bibinfo {author} {\bibfnamefont {L.}~\bibnamefont {Yu}},\ }\bibfield
  {title} {\bibinfo {title} {Kondo screening coexisting with ferromagnetic
  order as a possible ground state for kondo lattice systems},\ }\href
  {https://doi.org/10.1103/PhysRevB.81.094420} {\bibfield  {journal} {\bibinfo
  {journal} {Phys. Rev. B}\ }\textbf {\bibinfo {volume} {81}},\ \bibinfo
  {pages} {094420} (\bibinfo {year} {2010})}\BibitemShut {NoStop}%
\bibitem [{\citenamefont {Liu}\ \emph {et~al.}(2013)\citenamefont {Liu},
  \citenamefont {Zhang},\ and\ \citenamefont {Yu}}]{Liu2013}%
  \BibitemOpen
  \bibfield  {author} {\bibinfo {author} {\bibfnamefont {Y.}~\bibnamefont
  {Liu}}, \bibinfo {author} {\bibfnamefont {G.-M.}\ \bibnamefont {Zhang}},\
  and\ \bibinfo {author} {\bibfnamefont {L.}~\bibnamefont {Yu}},\ }\bibfield
  {title} {\bibinfo {title} {Weak ferromagnetism with the kondo screening
  effect in the kondo lattice systems},\ }\href
  {https://doi.org/10.1103/PhysRevB.87.134409} {\bibfield  {journal} {\bibinfo
  {journal} {Phys. Rev. B}\ }\textbf {\bibinfo {volume} {87}},\ \bibinfo
  {pages} {134409} (\bibinfo {year} {2013})}\BibitemShut {NoStop}%
\bibitem [{\citenamefont {Peters}\ \emph {et~al.}(2012)\citenamefont {Peters},
  \citenamefont {Kawakami},\ and\ \citenamefont {Pruschke}}]{Peters2012}%
  \BibitemOpen
  \bibfield  {author} {\bibinfo {author} {\bibfnamefont {R.}~\bibnamefont
  {Peters}}, \bibinfo {author} {\bibfnamefont {N.}~\bibnamefont {Kawakami}},\
  and\ \bibinfo {author} {\bibfnamefont {T.}~\bibnamefont {Pruschke}},\
  }\bibfield  {title} {\bibinfo {title} {Spin-selective kondo insulator:
  Cooperation of ferromagnetism and the kondo effect},\ }\href
  {https://doi.org/10.1103/PhysRevLett.108.086402} {\bibfield  {journal}
  {\bibinfo  {journal} {Phys. Rev. Lett.}\ }\textbf {\bibinfo {volume} {108}},\
  \bibinfo {pages} {086402} (\bibinfo {year} {2012})}\BibitemShut {NoStop}%
\bibitem [{\citenamefont {Peters}\ and\ \citenamefont
  {Kawakami}(2017)}]{Peters2017}%
  \BibitemOpen
  \bibfield  {author} {\bibinfo {author} {\bibfnamefont {R.}~\bibnamefont
  {Peters}}\ and\ \bibinfo {author} {\bibfnamefont {N.}~\bibnamefont
  {Kawakami}},\ }\bibfield  {title} {\bibinfo {title} {Competition of striped
  magnetic order and partial kondo screened state in the kondo lattice model},\
  }\href {https://doi.org/10.1103/PhysRevB.96.115158} {\bibfield  {journal}
  {\bibinfo  {journal} {Phys. Rev. B}\ }\textbf {\bibinfo {volume} {96}},\
  \bibinfo {pages} {115158} (\bibinfo {year} {2017})}\BibitemShut {NoStop}%
\bibitem [{\citenamefont {Gole\ifmmode~\check{z}\else \v{z}\fi{}}\ and\
  \citenamefont {\ifmmode~\check{Z}\else
  \v{Z}\fi{}itko}(2013)}]{PhysRevB.88.054431}%
  \BibitemOpen
  \bibfield  {author} {\bibinfo {author} {\bibfnamefont {D.}~\bibnamefont
  {Gole\ifmmode~\check{z}\else \v{z}\fi{}}}\ and\ \bibinfo {author}
  {\bibfnamefont {R.}~\bibnamefont {\ifmmode~\check{Z}\else \v{Z}\fi{}itko}},\
  }\bibfield  {title} {\bibinfo {title} {Lifshitz phase transitions in the
  ferromagnetic regime of the kondo lattice model},\ }\href
  {https://doi.org/10.1103/PhysRevB.88.054431} {\bibfield  {journal} {\bibinfo
  {journal} {Phys. Rev. B}\ }\textbf {\bibinfo {volume} {88}},\ \bibinfo
  {pages} {054431} (\bibinfo {year} {2013})}\BibitemShut {NoStop}%
\bibitem [{\citenamefont {Peters}\ and\ \citenamefont
  {Kawakami}(2012)}]{PhysRevB.86.165107}%
  \BibitemOpen
  \bibfield  {author} {\bibinfo {author} {\bibfnamefont {R.}~\bibnamefont
  {Peters}}\ and\ \bibinfo {author} {\bibfnamefont {N.}~\bibnamefont
  {Kawakami}},\ }\bibfield  {title} {\bibinfo {title} {Ferromagnetic state in
  the one-dimensional kondo lattice model},\ }\href
  {https://doi.org/10.1103/PhysRevB.86.165107} {\bibfield  {journal} {\bibinfo
  {journal} {Phys. Rev. B}\ }\textbf {\bibinfo {volume} {86}},\ \bibinfo
  {pages} {165107} (\bibinfo {year} {2012})}\BibitemShut {NoStop}%
\end{thebibliography}%

\end{document}